\documentclass[12pt,a4paper]{article}
\pdfoutput=1
\usepackage{jheppub}
\usepackage{amsmath}   
\usepackage{amsfonts}
\usepackage{mathtools}
\usepackage{amssymb}
\usepackage{caption}
\usepackage{subcaption}
\usepackage{bm}
\usepackage{cleveref}
\usepackage{dsfont}
\crefname{equation}{equation}{equations}
\Crefname{equation}{Equation}{Equations}% For beginning \Cref
\crefrangelabelformat{equation}{(#3#1#4--#5#2#6)}

\crefmultiformat{equation}{equations (#2#1#3}{, #2#1#3)}{#2#1#3}{#2#1#3}
\Crefmultiformat{equation}{Equations (#2#1#3}{, #2#1#3)}{#2#1#3}{#2#1#3}
\def\be{\begin{equation}}
\def\ee{\end{equation}}
\def\ba{\begin{array}}
\def\ea{\end{array}}
\newcommand{\bea}{\begin{eqnarray}}
\newcommand{\eea}{\end{eqnarray}}
\title{Chaos bound in Bershadsky-Polyakov theory}
\author{Justin R. David${}^{a}$, Timothy J.  Hollowood${}^{b}$, Surbhi Khetrapal${}^{c}$, S. Prem Kumar${}^{b}$}
\affiliation{${}^{(a)}$ Centre for High Energy Physics, Indian Institute of Science,\\
C. V. Raman Avenue, Bangalore 560012, India. \\
${}^{(b)}$ Department of Physics, Swansea University, \\
Singleton Park, Swansea SA2 8PP, UK. \\
${}^{(c)}$ Theoretische Natuurkunde, Vrije Universiteit Brussel (VUB) and \\
The International Solvay Institutes, Pleinlaan 2, B-1050 Brussels, Belgium}
\emailAdd{justin@iisc.ac.in, t.hollowood@swansea.ac.uk, surbhi.khetrapal@vub.be, s.p.kumar@swansea.ac.uk, }
\abstract{We consider  two dimensional conformal field theory (CFT) 
with large central charge $c$ in an excited state obtained by the insertion of an operator $\Phi$ with large dimension $\Delta_\Phi \sim O(c)$ at spatial infinities in the thermal state. We argue that correlation functions of light operators in such a state can be viewed as thermal correlators with a rescaled effective temperature.  
The effective temperature controls the growth of out-of-time order (OTO) correlators and results in a violation of the universal upper bound on the associated Lyapunov exponent when $\Delta_\Phi <0$ and the CFT is nonunitary. 
We present a specific realization of this situation in the holographic Chern-Simons  formulation of a CFT with ${W}^{(2)}_3$ symmetry also known as the Bershadsky-Polyakov algebra. 
We examine the precise correspondence between the semiclassical (large-$c$) representations of this algebra and the Chern-Simons formulation, and infer that the holographic CFT possesses a discretuum of degenerate ground states with negative conformal dimension $\Delta_\Phi\,=\,-\frac{c}{8}$.
Using the Wilson line prescription to compute entanglement entropy and OTO correlators in the holographic CFT undergoing a local quench,
we find the Lyapunov exponent $\lambda_L = \frac{4\pi}{\beta}$, violating the universal chaos bound. }

\begin{document}
\maketitle
\flushbottom

\section{Introduction}

It is well known that unitarity and causality   in conformal field theories (CFTs) have been used to  constrain
the allowed spectra  as well as to classify the  allowed theories. 
For example, positivity of the norm in two dimensional CFTs 
constrains both the conformal dimensions of operators and the central charge to be positive. In fact, demanding positive eigenvalues of the Kac matrix, enables  the classification of two dimensional CFTs with central charge $0< c<1$ and determination of the scaling dimensions of the primaries of these theories (see e.g. \cite{Ginsparg:1988ui}).  
Recently, constraints on  other physically interesting observables have been shown to be related to 
unitarity and causality. For example,  the fact that energy deposited in the detector of the conformal
collider  is positive \cite{ Hofman:2008ar}, or the average null energy condition,  has been shown to be related to unitarity and causality 
 \cite{Hartman:2015lfa, Hofman:2016awc, Hartman:2016lgu}. 
 The bound on the Lyapunov index which controls the growth of out-of-time ordered four-point functions,  
 has also been argued to be related to unitarity and causality \cite{mss}.

In  two dimensions,  it was was shown that  unitarity, or more specifically, the positivity of the eigenvalues of the Kac matrix  implies 
a bound on the spin-3 charge in a CFT with ${ W}_3$ 
symmetry \cite{Afkhami-Jeddi:2017idc}. In the large-$c$ limit the bound  is given by,
\begin{equation}\label{b2}
9 \left( \frac{\cal W}{c}\right)^2 \leq \frac{64}{5} \left( \frac{\Delta}{c} \right)^3 - \frac{2}{5} \left( 
\frac{\Delta}{c} \right)^2\,.
\end{equation}
Here ${\cal W}$ is the spin-3 charge and $\Delta$  the conformal dimension of the primary carrying this charge\footnote{We hold $ \frac{\cal W}{c},\frac{\Delta}{c} $ fixed as $c\rightarrow \infty$.}. 
At the same time  it was independently shown in \cite{David:2017eno} that  
one can arrive at a bound on the higher spin charge, in a holographic large-$c$ CFT with $W_3$ symmetry, by demanding 
that the jump in entanglement entropy during a quantum quench by an operator 
carrying spin-3 charge is always real:
\begin{equation}\label{b1}
\sqrt{\frac{ | {\cal W}| }{c}  }< \frac{2\Delta}{c}\,.
\end{equation}
It was also seen in \cite{David:2017eno}  
that the bound (\ref{b1}) automatically ensures that the chaos bound on the Lyapunov index for the
out-of-time ordered four-point function is obeyed
\footnote{In \cite{David:2017eno} the holographic ${\rm SL}(3) \times {\rm SL}(3)$ Chern-Simons Wilson line was used to compute the OTO correlator following a quench, confirming the arguments of \cite{Perlmutter:2016pkf} leading to a spin-three Lyapunov exponent $\lambda^{(3)} = 4\pi/\beta$. The general result for ${\rm SL}(N)$ higher spin gravity was shown to be $\lambda^{(N)} = 2\pi (N-1)/\beta$ in \cite{Narayan:2019ove}}.
Remarkably, if we restrict ourselves to the unitarity bound (\ref{b2}), 
then (\ref{b1}) is always satisfied.

\begin{figure}[h]
\center
\includegraphics[scale=0.35]{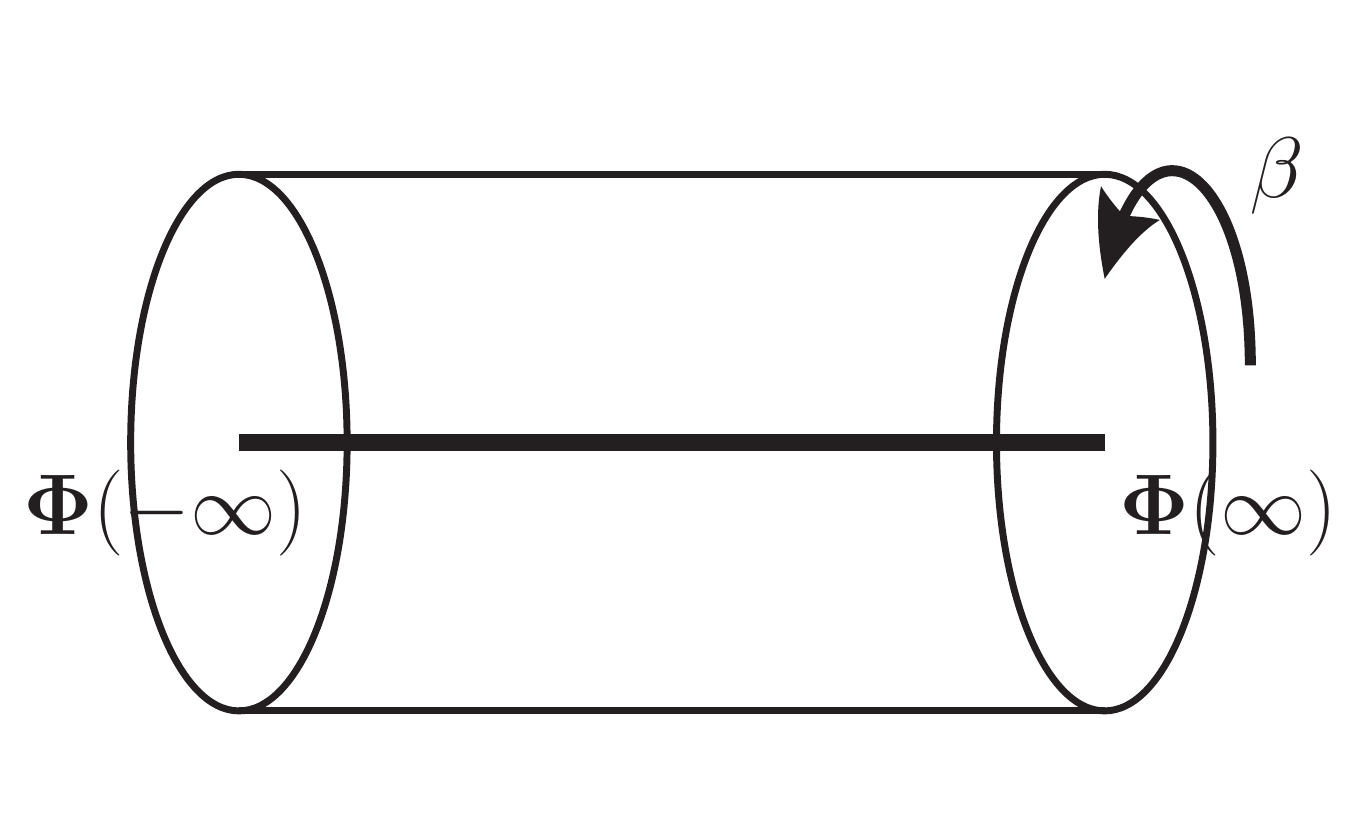}
\caption{\small {CFT at finite temperature $\beta^{-1}$ with heavy operators at spatial infinity.}}
\label{first}
\end{figure}

\begin{figure}[h]
\center
\includegraphics[scale=0.35]{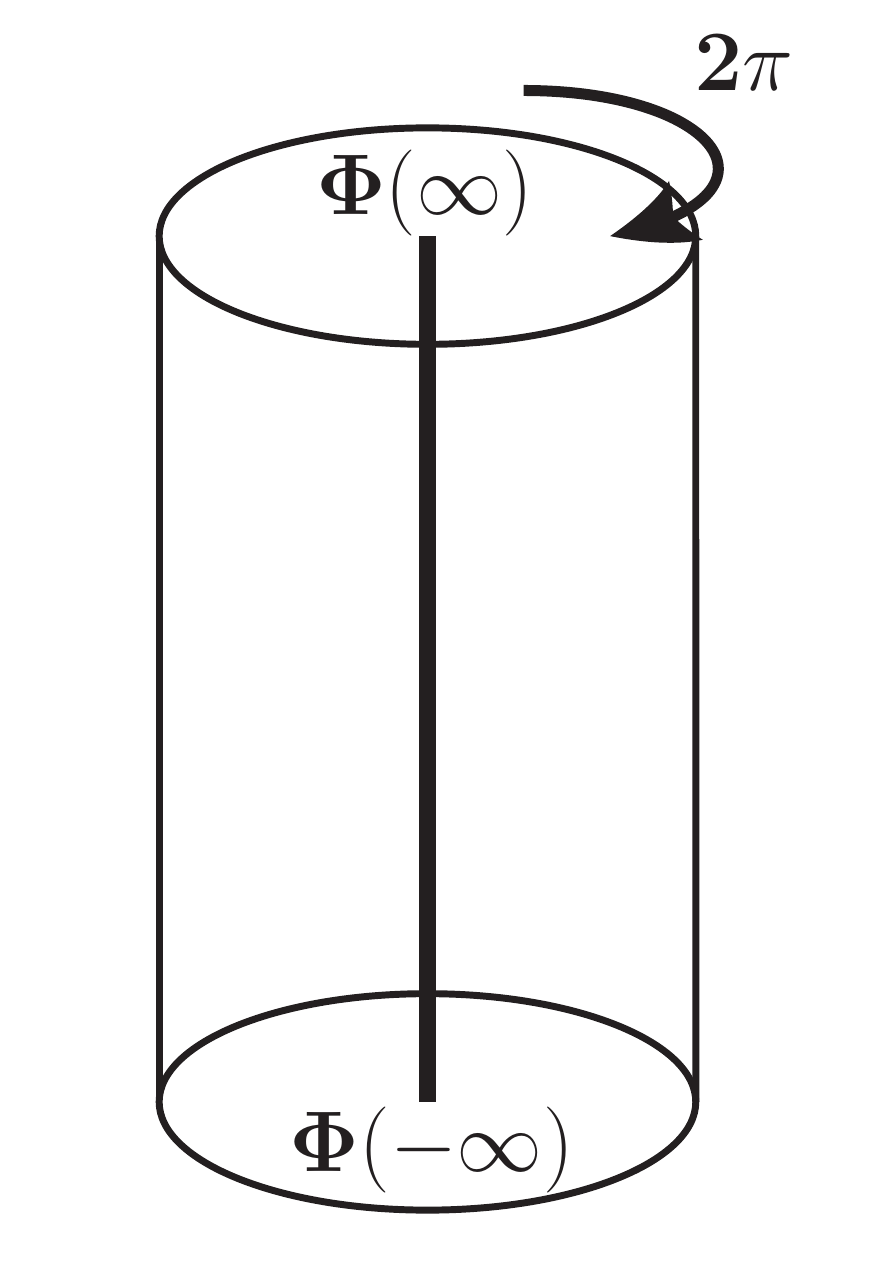}
\caption{\small CFT on a cylinder with heavy operators at temporal infinity}
\label{second}
\end{figure}

In this paper we would like to show that the chaos bound is also violated in certain holographic CFTs with (non-principal)  W-symmetries but with spin  $s\leq 2$ currents. The mechanism leading to violation of the bound is distinct from that   demonstrated in $W_N$ theories  with spin $s>2$ currents \cite{Afkhami-Jeddi:2017idc,  Perlmutter:2016pkf} . We will see that the violation here is due to the existence of a  nontrivial ground state (corresponding to an operator say, $\Phi$) with large, {\em negative} conformal dimension, $\Delta_{\Phi} <0$ and $|\Delta_\Phi| \sim O(c)$.  To understand this  we consider the setup shown in figure \ref{first}.
%In this paper we would like to show, amongst other things, that the chaos bound in 
%2d CFTs at large central charge is violated if there exists 
%an operator $\Phi$ with conformal dimension $\Delta_{\Phi} %<0$ and   $|\Delta_\Phi| \sim O(c)$ in its spectrum. 
%To do this we consider the setup shown in figure \ref{first}. 
Here we consider a conformal field theory at large central charge $c$ on an infinite 
line and at finite temperature $\beta^{-1}$. At the spatial infinities we insert an 
operator of conformal weights $ (\Delta_\Phi, \Delta_\Phi)$ such that $|\Delta_\Phi| \sim O(c)$. 
The more familiar situation in the literature is  when such an operator is 
inserted at temporal infinities on a cylinder of radius $2\pi$ as shown in figure \ref{second}. 
In this situation the   vacuum energy of the cylinder is shifted from its usual conformal value (summing over both holomorphic and antiholomorphic sectors),
\begin{equation}\label{vacen}
{\cal E}\,+\,\bar {\cal E}\, =\, -\frac{c}{12} \,\to\, - \frac{c}{12} \left( 1- \frac{24 \Delta_\Phi}{c}  \right) \,.
\end{equation}
Furthermore, the analysis of \cite{Asplund:2014coa,Fitzpatrick:2015zha} 
shows that correlators of light operators  inserted in the 
vacuum shown in figure \ref{second}  can be reproduced by considering the transformation to the plane
\begin{equation}
z\, =\, e^{ - i \alpha  \tilde w} , \qquad \alpha = \sqrt{ 1- \frac{24\Delta_\Phi}{c}}\,.
\end{equation}
Here $z$  is the coordinate in the plane and $\tilde w$ is the coordinate on the cylinder. 
It is  easy to see  the vacuum energy (\ref{vacen})  follows from the 
Schwarzian of the above 
transformation. 
We thus see that the cylinder acquires a deficit angle $2\pi (1- \alpha) $. Holographically, this
vacuum is dual to a conical defect created by a particle of mass $m = (\Delta_\Phi+ \bar\Delta_\Phi)/R$
where $R$ is the radius of AdS$_3$. 
The thermal state described in figure \ref{first}, though not very familiar, can be thought of as a Wick rotation 
of that in figure \ref{second}.  
The presence of the operator $\Phi$ shifts  the  energy density of the thermal state,
\begin{equation}\label{themen}
{\cal E}\,+\,\bar{\cal E}\, = \,\frac{c \pi^2}{3\beta^2}  \,\rightarrow \,\frac{c \pi^2}{3\beta^2} \left( 1- \frac{24 \Delta_\Phi}{c}  \right) \,.
\end{equation}
Note that we can obtain (\ref{vacen}) from (\ref{themen}) by the replacement $\beta \rightarrow  2i \pi $. 
We show that correlators of  operators with scaling dimensions much less than $\Delta_\Phi$
see an effective temperature different from $\beta$, given by 
\begin{equation}
\frac{1}{\beta_\Phi} \, =\, \frac{1}{\beta} \sqrt{  \left( 1- \frac{24 \Delta_\Phi}{c}  \right) }\,.
\end{equation}
This effective  temperature decreases when $\Delta_\Phi>0$ while it decreases when 
$\Delta_\Phi <0$.  In particular, we can argue that the rescaled temperature interpretation applies beyond one-point functions to include two-point functions of light operators whose correlators in the nontrivial vacuum, at large $c$, can be viewed as heavy-heavy-light-light correlators which can be evaluated as in \cite{Fitzpatrick:2014vua, Asplund:2014coa}. 
Extending the argument to include correlators of two additional heavy and two light operators in the nontrivial state (as required for computing EE/RE following a local quench),  we  conclude that the OTO four-point function of light operators  in the vacuum given in figure \ref{second} yield a shifted Lyapunov exponent,
\begin{equation}\label{lyapunovnew}
\lambda_L\,  =\, \frac{2\pi}{\beta} \sqrt{ 1- \frac{24\Delta_\Phi}{c} }\,.
\end{equation}
Thus the presence of a negative dimension operator  would violate the chaos bound of \cite{mss} which proposes $\lambda_L \leq 2\pi/\beta$. This is in turn  consistent with the fact that the bound is expected to be violated in non-unitary theories.

Do such situations,  as depicted in figures \ref{first} and \ref{second}, with $\Delta_\Phi <0$ arise in known conformal field theories ?
Vacua with negative scaling dimensions  or ground states which break scale invariance 
are known to exist in non-unitary CFTs.  It is 
 such a ground  state which is responsible for a  positive single interval
entanglement entropy in CFTs with $c<0$ like the Lee-Yang non-unitary 
minimal model at $c=-22/5$ \cite{Bianchini:2014uta}.
In this paper, we demonstrate that such examples can also arise holographically, i.e vacua with $\Delta_\Phi<0$ and $|\Delta_\Phi| \sim O(c)$  are realised in semiclassical holographic setups at large-$c$.  For this we examine the holographic realization of a CFT with a ${ W}_3^{(2)}$ symmetry. This algebra was  first considered by Polyakov \cite{Polyakov:1989dm} and Bershadsky \cite{Bershadsky:1990bg} and   
 is reviewed in some detail in 
section \ref{secw32}.  The algebra is generated by the stress tensor, two spin $3/2$ currents  and a $U(1)$ current, the latter at level $\kappa$.  The central charge of the algebra is related  to the $U(1)$ level $\kappa$ via, 
\be
  c\,=\,\frac{(2k-3)(3k-1)}{k-3}\ ,\qquad \kappa\,=\,-\frac{2k-3}3\,,\label{chat}
 \ee
 where $k$ is the level of the ${\mathfrak {sl}}(3)$ affine algebra from which the $W$-algebra arises by the procedure of Drinfel'd-Sokolov reduction. 
 Within the holographic framework it can be understood in terms of a  non-principal embedding of SL(2, ${\mathbb R}$) $\times$ SL(2, ${\mathbb R}$) within
${\rm SL}(3, {\mathbb R}) \times {\rm SL}(3, {\mathbb R})$ Chern-Simons  theory \cite{Ammon:2011nk, paper1} at level $k$.

 The algebra has  positive central charge which can be taken to infinity and therefore studied in the semiclassical 
 limit.  Since the level $\kappa$ of the $U(1)$ current algebra becomes negative, the semiclassical limit renders the associated theory nonunitary.  Nevertheless, the $W_3^{(2)}$ algebra has certain interesting nontrivial features.
 We first use the representation theory of the ${W}_3^{(2)}$ algebra to show that the 
theory admits a  one parameter set of degenerate ground states. 
The holographic Chern-Simons framework reproduces this ground state manifold and in addition fixes the conformal dimension of each of these states to be $\Delta_\Phi\, =\, - \frac{c}{8}$.  The individual states are distinguished by their spin-1 and spin-$\tfrac32$ charges. 
The value of the conformal dimension of the ground state and the resulting  non-trivial vacua of 
the kind shown in figures \ref{first} and \ref{second} follows from 
 the smoothness condition which demands that the holonomy of the Chern-Simons connection  be trivial 
along the spatial  or  temporal circles  at zero  and high temperatures respectively.

We probe aspects of  the excited thermal  state  admitted by the ${W}_3^{(2)}$ CFT (as in figure \ref{second}) in more detail. We evaluate the holographic entanglement entropy using the Wilson 
line prescription of \cite{deBoer:2013vca,Ammon:2013hba} 
and confirm that its  short distance expansion nontrivially matches that expected from heavy-heavy-light-light correlators  in a  thermal state excited by an operator of  dimension $\Delta_\Phi = - \frac{c}{8}$. 
The Wilson line correlator can also be used to compute the change in  EE  following a local quench which can be viewed as an infalling conical deficit state (carrying $U(1)$ and spin-$\tfrac32$ charges) in the $W_3^{(2)}$ black hole state. Following this correlator by analytic continuation into the Regge limit \cite{ Perlmutter:2016pkf} we obtain the OTO four point-function and show that the bound of \cite{mss} on the Lyapunov exponent is violated as would be expected in eq.\eqref{lyapunovnew} with $\Delta_\Phi = -\frac{c}{8}$.

The organization of the paper is as follows: in section 2, we lay out the general arguments for the appearance of a rescaled temperature in one-point functions and higher-point functions of light operators, in the presence of the nontrivial vacuum $|\Phi\rangle$. In section 3, we briefly review the holographic Chern-Simons realization of the CFT with $W_3^{(2)}$ symmetry. We also examine the representation theory of this algebra in the semiclassical limit and see how its nontrivial features are reproduced by the Chern-Simons framework. This allows the identification of the ground state conformal dimension as $\Delta_\Phi = -\frac{c}{8}$. Sections 4 and 5 deal with evaluation of holographic entanglement entropy using the Wilson line prescription.
In section 6 we discuss the calculation of the OTO correlator using the Wilson line EE in the presence of a local quench. We elaborate on various open ends and interesting questions for future work in section 7.

\section{Properties of a non-trivial vacuum in 2d CFT}
We will be interested in a CFT state obtained by the action of an $(h,\bar h)$ conformal primary ${\Phi}$ with conformal weights $h\,=\,\bar h\,=\,\Delta_\Phi$ on the conformal vacuum,
\be
{\Phi}\left|0\right\rangle\,=\,\left|\Phi\right\rangle\,.
\ee
At zero temperature, for the CFT on a spatial circle of length $\tilde L$, we may consider CFT ``in-out" amplitudes,
\be
\langle\Phi(-t_0)\rangle|\ldots\left|\Phi(t_0)\right\rangle\,,\qquad t_0\to\infty.
\ee
By modular invariance of the finite temperature CFT partition function, at high temperature $\beta \ll \tilde L$ we are then led to  consider the following density matrix: 
\begin{equation}
\rho_{\Phi }\, =\, {\Phi} (-\tilde L /2) e^{-\beta H}\ {\Phi(\tilde L/2 )}\,,\qquad \tilde L\to \infty\,,
\end{equation}
where it is important that the operator insertions  are kept at spatial infinity. The putative interpretation of such a state emerges upon using modular invariance to deduce the high temperature free energy. As is customary we  first normalize the spatial  length $\tilde L=2\pi$ and define $\tau=i\beta/(2\pi)$. Then modular invariance of the partition function,
 \be
 Z\left( -{\tau^{-1}}, - {\bar \tau}^{-1} \right) \,=\, Z(\tau, \bar \tau)\,, \qquad \tau\,=\,\frac{i\beta}{2\pi}\,,
\ee
constrains the free energy in the limit $\tau\to i0 $:
\bea
F\,=\,-\,\left.\beta^{-1}\,\ln Z\right|_{\tau\to i 0^+}\,\simeq\,\beta^{-1}\left[\frac{2\pi i}{\tau}\, \left(L_{0\,{\rm min}}-\tfrac{c}{24}\right)\,-\,\frac{2\pi i}{\bar \tau} \,\left(\bar L_{0\,{\rm min}}-\tfrac{c}{24}\right)\right]\,.
\eea
Here $L_{0\,{\rm min}}\,=\,\bar L_{0\,{\rm min}}\,=\,\Delta_\Phi$ is the conformal weight of the (zero temperature) ground state, so that
\be
F(\tau\to i0^+)\,=\, -\,\frac{\pi^2 c}{3\beta^2}\left(1\,-\,\frac{24\Delta_\Phi}{c}\right)\,.
\ee
At this stage, we could have two possible interpretations when compared with the result for the usual thermal state in a CFT with a conformal vacuum with $\Delta_\Phi\,=\,0$:
\begin{enumerate}
\item It is tempting to propose, following e.g. \cite{Bianchini:2014uta}, that the central charge is shifted to a new ``effective value" $c_{\rm eff}$,
\begin{equation}
c\, \rightarrow \,c_{\rm eff} \,=\, c\, -\, 24\Delta_\Phi\,.
\end{equation}
\item We will find evidence that favours an alternative interpretation, namely that the temperature of the theory  is shifted to a new  effective value,
\begin{equation} \label{defnewtemp}
\beta\, \rightarrow\,\beta_{\rm \Phi} \,=\, \frac{\beta}{ \sqrt {1 - \frac{24 \Delta_\Phi}{c} }}\,.
\end{equation}
\end{enumerate}
We are able to argue that it is the second interpretation which is consistent with the behaviour of two- and three-point correlation functions. Furthermore, while either interpretation could be adopted to explain  changes or shifts in one-point functions of observables, a shift in the central charge would  find its way into the stress tensor OPE. Since this would imply a modification of the short-distance properties, we disfavour this interpretation.  

\subsection{Correlators in the excited state}
\paragraph{One-point functions:} We now explain how the interpretation of a rescaled temperature appears via  correlation functions. Consider first the expectation value of the stress tensor in such a state,
\be
\langle T \rangle_{\beta,{\Phi}}\,\equiv\,\frac{\langle {\Phi } ( \infty) \,T_{ww} (w)\, {\Phi }( -\infty) \rangle_\beta}{
\langle {\Phi } ( \infty)\,  {\Phi}( -\infty) \rangle_\beta }\,.
\ee 
Here $w$ is the co-ordinate on the thermal cylinder related to the coordinate on the $z$-plane  as $z\,=\,e^{2\pi w/\beta}$. Since ${\Phi}$
is a primary, the expectation value can be deduced by application of the  
conformal Ward identity for the three-point function involving the stress tensor and two conformal primaries, 
\begin{eqnarray}
\langle T \rangle_{\beta,{\Phi}}
 \,= \,\left( \frac{2\pi }{\beta}\right) ^2 \Delta_\Phi \,-\, \frac{c\pi^2}{ 6 \beta^2}\,,
 \label{onepthighT}
\end{eqnarray}
where the second term is the usual thermal energy density which arises from the Schwarzian for the transformation from the plane to the thermal cylinder. 
Let us rewrite the expression for this expectation value more suggestively:
\begin{equation}
\langle T\rangle_{\beta, {\Phi}}  = 
- \frac{c \pi^2}{6 \beta^2} \left( 1- \frac{24 \Delta_\Phi}{c} \right) \,=\,- \frac{c \pi^2}{6 \beta^2_\Phi} \,.\label{stress1pt}
\end{equation}
As pointed out above, there are two possible ways to interpret the result when compared to the result for the usual thermal vacuum. 
Since the central charge can be measured by a local OPE which is unchanged from the usual one,  we choose the 
second option. Thus the effect of the presence of a nontrivial local operator in the thermal 
state is reflected by the change in the effective temperature according to \eqref{defnewtemp}.  In particular,  the effective temperature of the thermal state {\em decreases} if it is excited 
by an operator of {\em positive} conformal dimension, while it {\em increases} if excited by  a  {\em negative} dimension operator.  

The operator $\Phi$ we consider will be such that 
$\Delta_{\Phi} \sim O(c)$ in the large $c$ limit.  For later reference we can  also evaluate the entropy of this state. The total energy density is given by the sum of the expectation values of the stress tensor in the holomorphic and antiholomorphic sectors,
\begin{eqnarray}\label{enden}
{\cal E}_{\rm tot} \,=\,- \langle T_{ww} \rangle_\Phi\, -\, \langle T_{\bar w \bar w} \rangle_\Phi\,=\,\frac{c\pi^2 }{3\beta^2_\Phi } \,. 
%\left( 1- \frac{24\Delta_\Phi}{c} \right), 
\end{eqnarray}
We now examine what happens to the entropy of such a state.
The first principles approach, following Cardy \cite{Cardy:1986ie}, is to identify the density of states at high temperature as the saddle point of the inverse Laplace transform of the partition function,
\be
\varrho({\cal E})\,\simeq\,\int {d\tau} \,Z(\tau)\, e^{-2\pi i\tau{\cal E} }\,,\label{lapinv}
\ee
where we have only focussed on the holomorphic sector for clarity. Using modular invariance in the presence of the nontrivial ground state in the high temperature limit,
\be
Z(\tau)\Big|_{\tau \to i0^+} \simeq e^{-2\pi i\left(\Delta_\Phi-\tfrac{c}{24}\right)/\tau}\,,
\ee
we identify the saddle point of the integral \eqref{lapinv} and find the value of the density of states at the saddle point to yield the total entropy (summing over both chiral sectors),
\be
S_\Phi\,=\, \left.\ln \varrho \right|_{\rm saddle-pt} \,=\,\frac{2\pi^2 c}{3\beta}\,\left(1-\frac{24\Delta_\Phi }{c}\right)\,.
\label{sphi}
\ee
This is consistent with the usual thermodynamic relation $dS_\Phi\,=\, \beta d{\cal E}_{\rm tot}$ which leads to the same result. The form of the entropy does not immediately support the proposed temperature rescaling \eqref{defnewtemp} as the most natural interpretation.  For this we need to examine two- and higher-point correlation functions in the large-$c$ limit.

\paragraph{Two-point function in heavy state (HHLL):} We can now consider probing the excited heavy  state $|\Phi\rangle$ by two light operators of weights $h_L \sim O(1)$, so we are interested in the correlator,
\begin{equation} \label{2pths}
{\cal C}_{4}\,=\, 
\frac{ \langle {\Phi}(\infty)\, {\cal O}_L( w_1, \bar w_1) \,{\cal O}_L( w_2, \bar w_2) \,
{\Phi}(-\infty) \rangle_\beta }
{\langle {\Phi } ( \infty)  {\Phi }( -\infty) \rangle_\beta }\,,
\end{equation}
in a large-$c$ CFT.
Since the dimension of the light operators is parametrically smaller than 
that of the heavy operators and we take $c\rightarrow \infty$,  we can use 
the monodromy method \cite{Fitzpatrick:2014vua, Asplund:2014coa} to evaluate this correlator.  Before we do this  we can  first check what the temperature rescaling proposal \eqref{defnewtemp} would yield. According to this, we should simply obtain the thermal two-point function of an $(h_L, h_L)$ primary:
\begin{eqnarray}
&&\widetilde{\cal C}_{4} \,=\, \langle {\cal O}_L( w_1, \bar w_1) {\cal O}_L( w_2, \bar w_2) 
\rangle_{\beta_\Phi}\,=\,  
 \frac{\pi^{4h_L}}{ \left[\beta_\Phi^2\,\sinh\frac{\pi}{\beta_\Phi} ( w_1 - w_2) \, \sinh\frac{\pi}{\beta_\Phi } ( \bar w_1 - \bar w_2) \right]^{2h_L}}\nonumber\\
&& \beta_\Phi\,=\, \frac{\beta}{\sqrt{ 1- \frac{24\Delta_\Phi}{c} }}\,.
\end{eqnarray}
We now proceed to evaluate ${\cal C}_{4}$ as defined in (\ref{2pths}), by thinking of it as  the four-point function  involving two heavy states and 
two light states in the large $c$ limit.  Using the result for the conformal block
in this limit, dominated by the vacuum exchange, we obtain for the  
holomorphic part of the correlator on the $z$-plane,
\begin{eqnarray}\label{confblock}
 &&\left. \frac{\langle {\Phi }( \infty)\, {\cal O }_L ( z_1, \, \bar z_1 ) 
{\cal O }_L ( z_2,  \,\bar z_2) {\Phi}( 0 ) \rangle}{\langle {\Phi}( \infty) {\Phi}(0) \rangle}\right|_{\rm holomorphic} 
\\ \nonumber
&&\hspace{1.3in}\,=\, 
 \exp\left\{ - \tfrac{c \,\epsilon_L }{6} \left(
( 1- \alpha_\Phi) \log ( z_1 z_2) \,+\, 2 \log\frac{ ( z_1^{\alpha_\Phi} - z_2^{\alpha_\Phi})  }{\alpha_\Phi} 
\right) \right\}\,,
\end{eqnarray}
where 
\begin{equation}
\epsilon_L = \frac{6 h_L}{c}, \qquad \alpha_\Phi\, =\, \sqrt{ 1- \frac{24\Delta_\Phi}{c}} \,.
\end{equation}
This is  obtained by applying the 
monodromy method  in the HHLL limit, focussing only on the vacuum block \cite{Fitzpatrick:2014vua, deBoer:2014sna}. 
The anti-holomorphic contribution is identical to the above 
with the replacement $z\rightarrow \bar z$. Now we 
can find the correlator at finite temperature $\beta$, by mapping to the thermal cylinder:
\begin{equation}
z\,=\,\exp \left( \frac{2\pi w}{\beta} \right) \,.
\end{equation}
Using the transformation rule for the 
primaries we find
\begin{eqnarray}
{\cal C}_{4} \, =\, \widetilde{\cal C}_{4}\,,
\end{eqnarray}
demonstrating that  probing the thermal state at temperature $\beta$  excited
by a primary of dimension $\Delta_\Phi $,  by two light operators {\em in the large-$c$ limit}, 
is equivalent to evaluating the two-point function of the light operators 
in a theory at temperature $\beta_\Phi$ given by \eqref{defnewtemp}. 

\paragraph{Three-point function of light operators (HHLLL):} Let us see what happens when we probe the state with $3$ light operators. 
The correlator of interest is 
\begin{eqnarray}
{\cal C}_{5}\,=\, 
\frac{ \langle {\Phi}(\infty) \,{\cal O}_L( w_1, \bar w_1)\, {\cal O}_L( w_2, \bar w_2) \,
{\cal O}_L( w_3, \bar w_3) 
{\Phi}(-\infty) \rangle_\beta }
{\langle {\Phi } ( \infty)  {\Phi }( -\infty) \rangle_\beta }\,.
\end{eqnarray}
We again use the fact that the presence of the  excited state can be viewed as  the CFT at temperature $\beta^{ -1}_\Phi$. 
For the holomorphic part of the correlator, the result is then given by 
\begin{eqnarray}\label{5ptfn}
& & \left.\widetilde{\cal C}_{5}\right|_{\rm holomorphic} \,=\, \\ \nonumber
& &  \left(  \tfrac{\pi}{\beta_\Phi}\right) ^{3h_L} \frac{C_{LLL}}
{\left[ \sinh\frac{\pi}{\beta_\Phi} ( w_1 - w_2) 
 \sinh\frac{\pi}{\beta_\Phi} ( w_2 - w_3)
 \sinh\frac{\pi}{\beta_\Phi} ( w_3 - w_1) \right]^{h_L} } \\ \nonumber
 & &+ \left (\tfrac{\pi}{\beta_\Phi}\right )^{2h_L}\left ( \tfrac{2\pi}{\beta_\Phi} \right)^{h_L} 
 \frac{ \kappa }{ ( \sinh \frac{\pi}{\beta_\Phi}  ( w_1 -w_2) )^{2h_L}  }\,+ \, \left(\tfrac{\pi}{\beta_\Phi}\right )^{2h_L} \left( \tfrac{2\pi}{\beta_\Phi} \right)^{h_L} \,
 \frac{ \kappa }{ ( \sinh \frac{\pi}{\beta_\Phi}  ( w_2 -w_3) )^{2h_L} } \\\nonumber
&& +  \left(\tfrac{\pi}{\beta_\Phi} \right)^{2h_L} \left( \tfrac{2\pi}{\beta_\Phi} \right)^{h_L}
 \frac{ {\kappa} }{ ( \sinh \frac{\pi}{\beta_\Phi}  ( w_3 -w_1) )^{2h_L} }\,.
\end{eqnarray}
Here $C_{LLL}$ is the structure constant of the light operators. 
The last three terms are contributions from the one-point function 
of the operators in the thermal vacuum with the rescaled temperature $\beta_\Phi$, and $\kappa$ is the strength of the
one-point function. 
We can  evaluate the correlator ${\cal C}_{5}$ on the
 $z$-plane using the monodromy method and compare with the above \cite{Banerjee:2016qca, Alkalaev:2015lca}. 
 The monodromy method applies when 
 one of the operators is well separated from the other two. Accordingly, for operator insertions at $\{z_{i=1,2,3}\}$,  taking the limit of small $z_3$,  the answer on the plane is given by, 
 \begin{eqnarray}
 &&{\cal C}_{5}\,= \,
 \exp\left\{ - \frac{c \epsilon_L }{6} \left(
( 1- \alpha_\Phi) \log ( z_1 z_2)  + 2 \log\frac{ ( z_1^{\alpha_\Phi} - z_2^{\alpha_\Phi})  }{\alpha_\Phi} 
\right) \right\}\left (\frac{1}{z_3}\right)^{h_L}\,.
 \end{eqnarray}
The map from the plane to the cylinder at temperature $\beta$, $\{z_i\,=\,e^{2\pi w_i/\beta}\}$ finally yields,
 \begin{eqnarray}
&& {\cal C}_{5}  \,=\,\left( \frac{\pi }{\beta_\Phi}\right) ^{2h_L} \left( \frac{2\pi}{\beta} \right)^{h_L}\,
 \frac{1}{\left[\sinh \frac{\pi}{\beta_\Phi} ( w_1-w_2)\right]^{2 h_L} } \,.
 \end{eqnarray}
 Comparing $\widetilde {\cal C}_{5}$  in  \eqref{5ptfn} with 
 ${\cal C}_{5}$ we see that the two agree in the limit  $w_3\rightarrow -\infty$, if we make the identification,
 \begin{equation}
 \kappa\,= \, \alpha^{-h_L}_\Phi\,.
 \end{equation}
 In this limit the connected part drops out, and only the
 disconnected part in this channel contributes. 
 
\paragraph{The HHLLLL correlator :}  The six-point function involving four light operators in a heavy state  is the essential ingredient required
for understanding the time evolution of entanglement entropy in the presence of a local quantum quench. Analytic continuation of these  yields OTO correlators which are diagnostics of late time chaotic behaviour. On the thermal cylinder, the correlator of interest is
 \begin{eqnarray}\label{6ptfn}
{\cal C}_{6}\,=\, 
\frac{ \langle {\Phi}(\infty) \,{\cal O}_L( w_2, \bar w_2) \,{\cal O}_L( w_3, \bar w_3) \,
{\cal O}^\prime_L( w_4, \bar w_4) \,
{\cal O}^\prime_L( w_1, \bar w_1) \,
{\Phi}(-\infty) \rangle_\beta }
{\langle {\Phi } ( \infty) \, {\Phi}( -\infty) \rangle_\beta }\,.
\end{eqnarray}
The light operator  ${\cal O}_L$ is taken to have conformal dimension 
$h_L$ while the scaling dimension of the other light field ${\cal O}_L^\prime$ is $h_L^\prime$  with 
\be
\Delta_\Phi \gg h_L^\prime \gg h_L\,.
\ee
% Overall it is a 6-point function.  We also consider the weight $\Delta_L >>h_L$. 
% The order of limits we consider is $ \Delta>> \Delta_L >> h_L$. 
As before, we first view the correlator  as a four-point function  in the thermal state with  rescaled temperature $1/\beta_\Phi$, 
 \begin{eqnarray}
 \tilde {\cal C} _{6}\, =\, \langle {\cal O}_L( w_2, \bar w_2) {\cal O}_L( w_3, \bar w_3) 
 {\cal O}^\prime_L( w_4, \bar w_4) 
{\cal O}^\prime_L( w_1, \bar w_1) \rangle_{\beta_\Phi}\,.
 \end{eqnarray}
 Since $h_L^\prime \gg h_L$ we  
 can simply use the four-point function involving two heavy and two light (HHLL) operator insertions  at the temperature $1/\beta_\Phi$. 
 This is given by 
 \begin{eqnarray}
\tilde {\cal C}_{6} &&=\, \label{hhll6pt}\\\nonumber 
&&\left( \frac{\pi}{\beta_\Phi} \frac{1}{\sinh\frac{\pi}{\beta_\Phi} ( w_1 -w_4)}\right)^{2 h_L^\prime}
\left( \frac{\pi}{\beta_\Phi} \frac{1}{\sinh\frac{\pi}{\beta_\Phi} ( w_2 -w_3)}\right)^{2 h_L} 
\left( \frac{ \alpha' ( 1-x )} { (1-x^{\alpha' }) x^{\frac{1-\alpha'}{2} } } \right)^{2h_L}  \end{eqnarray}
 where
 \begin{equation}
 \alpha' = \sqrt{ 1 \,-\, \frac{ 24h_L^\prime}{c} }, 
 \qquad x =  
 \frac{\sinh\frac{\pi}{\beta_\Phi} ( w_1-w_2) \sinh\frac{\pi}{\beta_\Phi} ( w_3-w_4) }
 { \sinh\frac{\pi}{\beta_\Phi} ( w_1-w_3) \sinh\frac{\pi}{\beta_\Phi} ( w_2 - w_4) }\,.\label{defalphax}
 \end{equation}
 Using our observations  on the four-point and five-point functions, we claim that  when $\Delta_\Phi \gg  h_L^\prime \gg h_L$, and in the large $c$ limit, 
 the six point function in (\ref{6ptfn}) is given by $\tilde {\cal C}_{6}$: 
 \begin{equation}\label{efftemp}
 {\cal C}_{6}\Big|_{\Delta_\Phi \gg h_L^\prime \gg h_L\,,\, c\rightarrow \infty}\, =\, \tilde {\cal C}_{6}\,.
 \end{equation}
It is easy to check this relation when the insertions of the same light operator
come close together.
Consider the correlator ${\cal C}_6$ in the limit  $w_2\rightarrow w_3$ and examine it on the plane  in the identity block. Applying the monodromy method and from the 
analysis of \cite{Fitzpatrick:2014vua, Asplund:2014coa}, we know that the correlator factorises into four-point functions 
as 
\begin{eqnarray}
{\cal C}_6|_{z_2 \rightarrow z_3}  = 
 \frac{\langle {\Phi }( \infty) {\cal O }_L ( z_2, \bar z_2 ) 
{\cal O }_L ( z_3, \bar z_3) {\Phi}( 0 ) \rangle}{\langle {\Phi}( \infty) {\Phi}(0)\rangle }
 \frac{\langle {\Phi }( \infty) {\cal O }^{\prime}_L ( z_1, \bar z_1 ) 
{\cal O }^{\prime}_L ( z_4, \bar z_4) {\Phi}( 0 ) \rangle}{\langle {\Phi }( \infty) {\Phi}(0)\rangle}\,.\qquad
\end{eqnarray}
The result for each of these four-point functions on the plane yields, 
\begin{eqnarray}
{\cal C}_6\Big|_{z_2 \rightarrow z_3}   &=& 
 \exp\left\{ - h_L\left(
( 1- {\alpha_\Phi}) \log ( z_2 z_3)  + 2 \log\frac{ ( z_2^{\alpha_\Phi} \,-\, z_3^{\alpha_\Phi})  }{\alpha} 
\right) \right\}  \\ \nonumber
& \times &
\exp\left\{ - h_L^\prime \left(
( 1- \alpha_\Phi) \log ( z_1 z_4)  + 2 \log\frac{ ( z_1^{\alpha_\Phi }\,- \,z_4^{\alpha_\Phi})  }{\alpha_\Phi} 
\right) \right\}\,.
\end{eqnarray}
Now transforming to the thermal cylinder with periodicity $\beta$, using $z\,=\,e^{2\pi w/\beta}$ we obtain,
\begin{eqnarray}
{\cal C}_{6}|_{w_2 \rightarrow w_3}  &=& 
\left( \frac{\pi}{\beta_\Phi} \frac{1}{\sinh\frac{\pi}{\beta_\Phi} ( w_1 -w_4)}\right)^{2h_L^\prime}
\left( \frac{\pi}{\beta_\Phi} \frac{1}{\sinh\frac{\pi}{\beta_\Phi} ( w_2 -w_3)}\right)^{2 h_L} \,,
\end{eqnarray}
which matches the proposed result:
\begin{eqnarray}
\widetilde{\cal C}_{ 6}\Big|_ {w_2 \rightarrow w_3 }\, =\, {\cal C}_{6}\Big|_{w_2 \rightarrow w_3}\,.
\end{eqnarray}
We take  this and the agreement of the four- and five-point functions in various factorization channels, as evidence for the relation 
proposed in eq.\eqref{efftemp}. 
\\

In our arguments above, we have focussed attention on correlation functions of light operators on the plane, in the presence of insertions of  the heavy state $\Phi$, and then transformed the result to the thermal cylinder at temperature $\beta^{-1}$. We then see the rescaled effective temperature  $\beta_\Phi^{-1}$ emerge in the process.  It is useful to motivate this by following a slightly different line of reasoning as done below.

Consider the  four-point correlator of two heavy and two light operators (HHLL)  at large-$c$ and some finite temperature $\hat \beta^{-1}$ . 
For the heavy operators with dimensions $h_L^\prime$ and the light operators  with dimension $h_L$, the HHLL correlator has the standard form:
\begin{eqnarray}\label{hhllbeta}
{\cal C}_4(\hat\beta) = 
\left( \frac{\pi}{\hat \beta} \frac{1}{\sinh\frac{\pi}{\hat \beta} ( w_1 -w_4)}\right)^{2h_L^\prime}
\left( \frac{\pi}{\hat \beta} \frac{1}{\sinh\frac{\pi}{\hat \beta} ( w_2 -w_3)}\right)^{2 h_L} 
\left( \frac{ \alpha' ( 1-x )} { (1-x^{\alpha' }) x^{\frac{1-\alpha'}{2} } } \right)^{2h_L} \nonumber\,,
\\
\end{eqnarray}
with the cross-ratio at temperature $\hat \beta^{-1}$,
\begin{equation}
x\,=\,  
 \frac{\sinh\frac{\pi}{\hat \beta} ( w_1-w_2) \sinh\frac{\pi}{\hat \beta} ( w_3-w_4) }
 { \sinh\frac{\pi}{\hat \beta} ( w_1-w_3) \sinh\frac{\pi}{\hat \beta} ( w_2 - w_4) }\,,
\end{equation}
and $\alpha^\prime$ as given in \eqref{defalphax}.
Now, the thermal state at temperature $1/\hat \beta$ in the large-$c$ theory can be thought of as a heavy state 
with effective dimension $\Delta_{\rm BH} > \frac{c}{24}$ such that,
\begin{equation}
\frac{1}{\hat \beta}\, =\, \frac{1}{2\pi} \,\sqrt{ \frac{24\Delta_{\rm BH}}{c} - 1}\, \equiv\, i \frac{\alpha}{2\pi}\,.
\end{equation}
We can then view the thermal  correlator as a four-point function evaluated
in the black hole heavy state $|\Phi\rangle$ (in the large $c$ limit), which is effectively a  {\em six}-point function at zero temperature. 
With this viewpoint we convert the correlator in (\ref{hhllbeta})  to the plane by the  conformal transformation 
\begin{equation}
z \,= \,e^{i w} \,,
\end{equation}
where $w$ has periodicity $2\pi$:
\begin{eqnarray}\label{6ptfn2}
{\cal C}^\prime_{6pt} &= &
\frac{ \langle {\Phi}(\infty) {\cal O}_L( z_2, \bar z_2) {\cal O}_L( z_3, \bar z_3) 
{\cal O}^\prime_L( z_4, \bar z_4) 
{\cal O}^\prime_L( z_1, \bar z_1) 
{\Phi}(0) \rangle }
{\langle {\Phi } ( \infty)  {\Phi}( 0) \rangle } \\ \nonumber \\ \nonumber
&=& 
\left( \frac{\alpha}{ (  z_2^\alpha - z_3^\alpha) ( z_2 z_3)^{\frac{1-\alpha}{2}} } \right)^{2h_L}
\left( \frac{\alpha}{ (  z_4^\alpha - z_1^\alpha) ( z_4 z_1)^{\frac{1-\alpha}{2}} } \right)^{2 h_L^\prime} 
\left( \frac{ \alpha' ( 1-x )} { (1-x^{\alpha' }) x^{\frac{1-\alpha'}{2} } } \right)^{2h_L}\,, 
\end{eqnarray}
and $x$ is the cross-ratio expressed on the plane,
\begin{equation}
x = \frac{ ( z_1^\alpha - z_2^\alpha) ( z_3^\alpha - z_4^\alpha) }
{(z_1^\alpha -z_3^\alpha)( z_2^\alpha - z_4^\alpha)}\,.\label{xplane}
\end{equation}
We may readily get from this correlator (\ref{6ptfn2})  on the plane to a thermal correlation function at any temperature $\beta ^{-1}$  by using the exponential map $z = e^{2\pi \tilde w/\beta}$.  The procedure yields  $\tilde {\cal C}_{6}$ with rescaled temperature as discussed above\footnote{While this work was being prepared, the preprint \cite{Anous:2019yku} appeared which has  overlap with the present discussion, particularly with regard to OTO correlators in the presence of a heavy state $\Phi$, and the interpretation in terms of an effective temperature $\beta_\Phi$. Crucial to this is the branch point  at $x=\infty$ (or equivalently $x=0$), around which $x$ must be rotated anticlockwise, $x\to x \,e^{2\pi i }$, in order to get the out-of-time ordering, and subsequently take the limit  $x\to 1$ to approach the Regge limit for chaos \cite{Perlmutter:2016pkf}.  In this limit, ignoring the two overall factors dictated only by the scaling dimensions of the operators $O_L, O_L^\prime$, the key dependence on the cross ratio becomes,
\be
\ln\tilde{\cal C}_6\big|_{x\to 1}\propto-2h_L\ln\left(1 \,+\,\frac{24i\pi h_L^\prime}{c(1-x)}\right)\,.
\ee
This expression translates in the real time setting into an exponential growth  of the OTO correlator, from which the Lyapunov exponent is read off. To obtain this final form, we have assumed a double scaling limit with $h_L^\prime \sim |1-x| \ll1$. On the plane the cross-ratio $x$ is given by eq.\eqref{xplane} with $\alpha=\sqrt{1-24 \Delta_\Phi/c}$. In \cite{Anous:2019yku}, a six-point correlator with two heavy insertions was deduced in the large-$c$ theory with no heirarchy between $h_L$ and $h_L^\prime$ i.e. $\Delta_\Phi\gg h_L^\prime, h_L$. Focussing attention on eq.(4.3) of that paper, after translating their insertion points in terms of ours, we can express their result as a function of our cross-ratio $x$ ( as in eq.\eqref{xplane}),
\be
{\cal C}_6(x) \sim \exp\left(-\tfrac{2 h_L h_L^\prime}{c}(1-x)^2\,{}_2F_1(2,2,4;1-x)\right)\,,
\ee
where we have again ignored the overall contributions proportional to the two-point functions of the operators $O_L, O_L^\prime$.  The hypergeometric function has a branch cut along $x\in\{0, -\infty\}$. The rotation by a $2\pi$-phase around the branch-point at $x=0$  shifts the hypergeometric function, which yields the dominant contribution to the OTO in the Regge limit,
\be
{\cal C}_6\big |_{x\to 1}\sim \lim_{x\to 1}\exp\left(-24\pi i\tfrac{h_L h_L^\prime}{c}(1-x)^2\,{}_2F_1(2,2,1; x)\right)\,=\,\exp\left(-2h_L \tfrac{24i\pi h_L^\prime}{c(1-x)}\right)\,,
\ee
which agrees with the limit obtained from our expression for $\tilde {\cal C}_6$ when formally expanded in powers of $h_L^\prime$ to lowest order.
  }.

 \subsection{ R\'{e}nyi entropy in an excited thermal state}
 \label{sec:renyi}
 
 We can  use the result for the two-point function of light 
 operators in the excited heavy state at finite temperature to evaluate 
R\'{e}nyi entropies.  The role of light operators is now played by the twist fields $\left(\sigma_n,\bar\sigma_n\right)$ used to implement the replica trick, and whose conformal dimension we take to be of the form \cite{Calabrese:2004eu}:
 \begin{equation}
 h_\sigma \,=\,  \frac{c}{24} \left( n - \frac{1}{n} \right)\,,
 \end{equation}
 with $n$ the number of replicas.
 First we look at the zero temperature correlator in the presence of the heavy state $|\Phi\rangle$ for the CFT on a circle:
 \begin{equation}\label{eecor}
  \left.{\cal C}_{4}\,\right|_{{\mathbb R}_t\times S^1}\, =\, \frac{ \langle {\Phi}(t\to \infty)\, \,\sigma_n (L)\,  \bar{\sigma}_n (0)\, \,{\Phi}(t\to -\infty)
  \rangle}
  {\langle{\Phi}(t\to\infty)\,\, {\Phi}(t\to -\infty) \rangle}\,.
  \end{equation}
Applying the conformal map from the cylinder to the plane,
 \begin{equation}
 z \,=\, \exp (2 \pi  i w) \,,
 \end{equation}
from \eqref{confblock} we arrive at the R\'enyi entropy,
 \begin{eqnarray}
 S_n \,=\, \frac{1}{1-n}  \left. \log {\cal C}_{4}\, \right|_{{\mathbb R}_t\times S^1} \, =\, c_p\frac{n+1}{6n}  \,
 \log \left(\frac{\sin ( \pi \alpha_\Phi L ) }{\pi \alpha_\Phi} 
  \right)\,, 
 \end{eqnarray}
 which matches the expected R\'{e}nyi entropy in the presence of a heavy operator or conical defect state.  For the theory at finite temperature then we  use the exponential map from the Euclidean thermal  cylinder to the plane, $z\,=\,\exp(2\pi w/\beta)$, and the 
R\'{e}nyi entropy is given by the replacement $L\to - i L/\beta$,
 \begin{eqnarray} \label{renyi}
 S_n (\beta)\,=\,\frac{1}{1-n}\, \log  {\cal C}_{4}(\beta)\, =\, \frac{n+1}{6n} c \log\left[
 \frac{\beta}{\pi \alpha_\Phi} 
 \sinh \left(\alpha _\Phi  \frac{\pi L}{\beta}   \right)  \right]\,.
 \end{eqnarray}
  This agrees with  holographic calculations of EE in a conical deficit state, and is non-trivial since it only relies on the vacuum block contribution to the HHLL correlator.  A short distance expansion yields a cross-check of the result.  Consider, under general assumptions  discussed below, the OPE of the twist-antitwist fields,
    \begin{equation} \label{opetwi}
  \sigma_n(L, L) \bar\sigma_n(0,0)\, \sim \,\frac{1}{L^{4h_\sigma}} \left( 
  1 \,+\, 2 \frac{h_{\sigma}}{c}\, L^2\, \left ( T_{ww} (0) \,+\, \bar T_{\bar w \bar w} ( 0) \right) \, + \cdots 
  \right) \,.
  \end{equation}
 Substituting this into the four-point function (\ref{eecor}),
and  using the stress tensor one-point function  (\ref{enden}) in the excited state, we find the short distance expansion of the R\'enyi entropy,
\begin{equation}
\lim_{L\to 0} S_n(\beta)\, = \,c\frac{n+1}{6n}\left(\log L \,+\, L^2 \frac{ \pi^2}{6 \beta^2} \left( 1 - \frac{24\Delta_\Phi}{c}\right)\,+ \cdots 
 \right) \,.
 \end{equation}
This coincides with the short distance expansion of 
(\ref{renyi}). 

Let us now briefly  discuss the validity of the OPE in eq.(\ref{opetwi}). 
\begin{enumerate}
\item  We are assuming that no primaries appear on the right hand side of \eqref{opetwi}. Note that we can derive the structure constant $\langle \sigma_n \bar\sigma_n {\cal O} \rangle $
for any primary field ${\cal O}$ using the uniformization map. Since primaries do not yield
Schwarzian-like shifts under conformal transformations (and assuming no one-point functions for ${\cal O}$) the relevant structure constant vanishes. 
\item Thus the only non-trivial contribution to the twist-antitwist OPE can arise from a quasi-primary field like 
the stress tensor, its composites or composites of other primary fields. 
\item  The fact that the leading term in the OPE is the identity operator assumes that there are no composite fields 
with negative conformal dimension. In particular, while we will allow for  
a  primary field  $\Phi$ with negative conformal dimension  $\Delta_\Phi <0$, we assume that there are no other quasi-primary fields  with negative dimensions in the theory. 
\item We have also assumed that the there are no quasi-primaries  with dimensions 
$0<\Delta\leq 2$ other than the stress tensor. 
Note that for a $U(1)$ current $J$, the composite $J^2$  is a quasi-primary of dimension $2$, but 
for the moment we assume that the theory does not admit such a current. 
\end{enumerate}
These comments justify the OPE in (\ref{opetwi}). 

We will be examining a situation in holography 
in which the vacuum not only has negative conformal dimension, but also carries
$U(1)$ charge. In this situation the holographic result admits a short distance 
expansion of the type in eq.\eqref{opetwi} accompanied by a contribution from $J^2$.

 \subsection{Local quench and OTO correlator}
 The four-point function (HHLL) in a black hole heavy state  (effectively the six-point correlator) can be used to compute the time evolution of entanglement entropy following a local quench \cite{Caputa:2014eta, Caputa:2015waa} i.e. a finite energy perturbation by a local operator. We examine the correlator ${\cal C}_{6}(\beta)$ with  the heavy operator ${\cal O}_L^\prime$ inserted at the locations $(w_1,\bar w_1)$ and $(w_4,\bar w_4)$, and twist field insertions at $(w_2, \bar w_2)$  and $(w_3, \bar w_3)$ which are lightcone coordinates of the end points of the interval of interest,
 \begin{eqnarray} \label{OTOcoords}
&w_1=-i\epsilon\,,\quad\qquad  &\bar w_1 = +i \epsilon\,, \\ \nonumber
 & w_2 = \ell_1 - t \,, \qquad &\bar w_2 = \ell_1 + t\,, \\ \nonumber
 &w_3 = \ell_2 - t\,, \qquad &\bar w_3 = \ell_2 + t\,,\\ \nonumber
 &w_4 = i \epsilon\,,\quad \,\,\qquad &\bar w_4 = - i \epsilon\,.
 \end{eqnarray}
 The parameter $\epsilon$ smears the excitation over a finite size pulse.
 We then use ${\cal C}_6(\beta)$ (eq.\eqref{hhll6pt}) to compute the change in entanglement entropy as a function of time (see e.g. \cite{David:2016pzn, David:2017eno}) where the state is thermal with insertions of the operator $\Phi$ at spatial infinity on the thermal cylinder,  the lightest operators being the twist fields ($h_L\,=\, h_\sigma$):
 \begin{eqnarray}
 S_{\rm EE} &=& \lim_{n\to1 } \frac{2}{1-n}  \ln \left |  {\cal C}_{6}  
 \left( \frac{\pi}{\beta_\Phi} \frac{1}{ \sinh\frac{\pi}{\beta_\Phi} ( w_1 - w_4) } \right)^{-2 h_L^\prime} 
 \right|\\ \nonumber\\ \nonumber
 &=& \frac{c}{6} \ln\left\{ \frac{\beta^2_\Phi}{\pi^2 \alpha^{'\,2} }
 \sinh^2 \frac{\pi}{\beta_\Phi} ( \ell_2 - \ell_1) \frac{ \left[ 1 - ( 1-z)^{\alpha^\prime}\right] \left[1- ( 1-\bar z)^{\alpha^\prime} \right]}{
 ( 1- z)^{\frac{\alpha^\prime-1}{2}}  ( 1- \bar z)^{\frac{\alpha^\prime-1}{2}}  z\bar z} \right\}\,.
 \end{eqnarray}
 The cross-ratio $z$ with rescaled temperature is given by,
 \begin{equation}
 z= \frac{( z_2 -z_3) ( z_1 - z_4)}{(z_2 - z_1)(z_3-z_4) }, \qquad 
 z_i = \exp\left( \tfrac{2\pi }{\beta_\Phi} \,w_i \right)\,. 
 \end{equation}
 The cross ratios $z$ is related to the previously introduced variable $x$ (see \eqref{defalphax}) as $x=1/(1-z)$.
% Note that these are identical to the situation discussed earlier, except that 
% we have $\beta\rightarrow \beta'$. 
Now we 
 scale the dimension of the operator ${\cal O}_L^\prime$  to be such that the 
 energy carried by the pulse is fixed in the limit of small width $\epsilon$ \cite{David:2017eno}. This requires $h_L^\prime$ to scale with $\epsilon$, while also allowing it to scale with $c$ in the large-$c$ limit. Thus
 \begin{equation}
 \frac{h_L^\prime}{c}\, =\, \frac{E}{\pi c} \epsilon \ll 1, \qquad \epsilon \ll \beta_\Phi\,,
 \end{equation}
 with $\frac{E}{\pi c} $ {\em fixed} in the large $c$ limit. 
 Then performing a small width expansion we obtain the time dependence of the entanglement entropy of the interval. When the pulse induced by the local quench enters the  interval $S_{\rm EE}$ grows, reaches a maximum and subsequently decreases as the pulse exits the interval:
 \begin{equation}
 \frac{6}{c} \left.  \Delta S_{\rm EE} \right|_{\ell_2 >t>t_1} \,=\, 
 \ln\left[ 1 + 12 \left( \frac{\beta_\Phi E}{\pi c}  {\cal Z}_{\ell_1 \ell_2}^{-1} ( t) \right)   \right]
 + O(\epsilon) 
 \end{equation}
 where
 \begin{equation}
  {\cal Z}_{\ell_1 \ell_2} ( t) \, =\,  \frac{\sinh\frac{\pi}{\beta_\Phi} ( \ell_2 - \ell_1) }
  {\sinh\frac{\pi}{\beta_\Phi} ( t - \ell_1) \sinh\frac{\pi}{\beta_\Phi} ( \ell_2 - t) }\,.
  \end{equation}
  The height of the plateau in the jump $\Delta S_{\rm EE}$ can be obtained easily in the limit of large interval length: 
  \begin{equation}
  \left. \Delta S_{\rm EE}\right|_{\ell_2\gg t \gg \ell_1}\,=\, 
  \frac{c}{6 }\ln \left( 1 + \frac{ 6 \beta_\Phi E}{\pi c}  \right)\,  =\, \frac{c}{6} \ln \left( 
  1+ \frac{2E}{s_{\Phi}} \sqrt{1 - \frac{24\Delta_\Phi}{c} } \right) \,.
  \end{equation}
 Here $s_{\Phi}$ is the thermal entropy density in the excited state (with insertion of the operator $\Phi$ at spatial infinity) following from  \eqref{sphi},
  \begin{equation}
  s_\Phi\,=\, \frac{S_{\Phi}}{2\pi}\,.
  \end{equation}
  The correlator used to obtain the EE following the quench also yields   an OTO correlation function upon appropriate analytic continuation \cite{David:2017eno} into the second sheet as a function of the cross-ratio $z$, and taking the so-called Regge limit \cite{Perlmutter:2016pkf} where the cross ratio $z$ is small. The OTO correlator in the excited thermal state is then (taking $t\gg \ell_2\gg \ell_1$),
 \begin{eqnarray}
 {\cal C}_{\rm OTO}\,=\,\left[  \frac{\beta^{2}_\Phi}{\pi^2} \,\sinh^2 \frac{\pi}{\beta_\Phi} ( \ell_2 - \ell_1) 
 \left( 1 \,-\,  \frac{6\beta_\Phi E}{\pi c} \,e^{\frac{2\pi }{\beta_\Phi} ( t + i \epsilon_1 - \ell_2 ) } \right) 
 \right]^{- 2 h_\sigma}\,.
 \end{eqnarray}
 Here an imaginary part $i\epsilon_1$ is added to $t$ which  ensures that the 
 correlator is evaluated on the second sheet for $t >(\ell_1 + \ell_2) /2$. The exponential
 decay of the correlator is governed by the Lyapunov exponent, 
 \begin{equation}
 \lambda_L \,= \,\frac{2\pi}{\beta_\Phi}\, =\, \frac{2\pi}{\beta} \sqrt{ 1- \frac{24\Delta_\Phi}{c}}\,.
 \end{equation}
 Therefore, with $\Delta_\Phi > 0$, the Lyapunov exponent is lower than the  
 Maldacena-Shenker-Stanford bound \cite{mss}, whereas the bound is violated when 
 when $\Delta_\Phi < 0$. 
 
 Hence we have shown if  a non-unitary theory  at large $c$ admits 
 a state of large negative  dimension, then the OTO correlator in that state has an associated  Lyapunov 
 exponent which violates the bound.  Below we will 
 consider the holographic (large-$c$) realization of a CFT with  $W_{3}^{(2)}$ symmetry, also known as the  Polyakov-Bershadsky algebra,
 which exhibits these features alongside other accompanying novel aspects.

 \section{$W_3^{(2)}$ symmetry and holography}
 \label{secw32}
Gravity in asymptotically AdS$_3$ spacetimes can  naturally be formulated in terms of Chern-Simons theory with ${\rm SL}(2,{\mathbb R})\times {\rm SL}(2,{\mathbb R})$ gauge symmetry \cite{Achucarro:1987vz, Witten:1988hc}. Higher spin symmetries can be incorporated by extending the gauge group and in particular, ${\rm SL}(N,{\mathbb R})\times {\rm SL}(N,{\mathbb R})$ Chern-Simons theory yields gravity and a (finite) tower of higher spin gauge fields on AdS$_3$.  In this situation, the asymptotic symmetry algebra is a  ${W}$-algebra. The ${W}$-symmetry depends on how gravity (as an ${\mathfrak{sl}(2)}$ subalgebra) is embedded in ${\rm SL}(N,{\mathbb R})$. The principal embedding yields a ${W}_N$ symmetry in the boundary CFT generated by the stress tensor and $(N-1)$ higher spin currents \cite{Henneaux:2010xg, Campoleoni:2010zq}. Other non-principal embeddings give rise to ${W}$-symmetries with half-integer spin currents. In the simplest situation with $N=3$, a black hole solution carrying spin three charge was found in \cite{Gutperle:2011kf, Ammon:2011nk, Ammon:2012wc} by turning on a chemical potential deformation for the associated current. The corresponding irrelevant deformation yields flat Chern-Simons connections that appear to interpolate between two different asymptotic symmetric algebras, namely the standard ${W}_3$ symmetry (in the IR) generated by the stress tensor and the spin three current, and a different (UV) ${W}_3^{(2)}$ symmetry algebra (from the `diagonal' embedding) generated by the stress tensor, two spin-$\frac{3}{2}$ currents and a $U(1)$ current.
\begin{figure}[h]
\begin{center}
\includegraphics[width=2.2in]{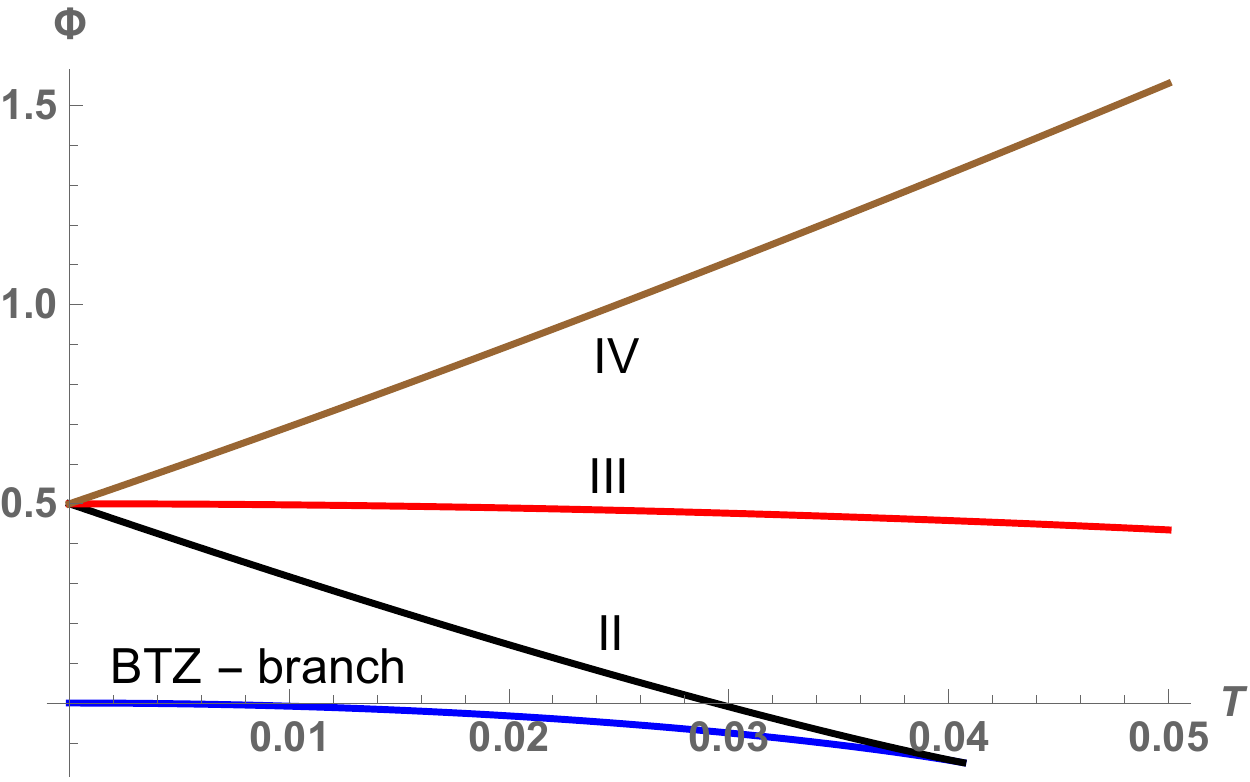}
%\hspace{1.0in}
%\includegraphics[width=2.2in]{d3spike2}
\end{center}
\caption{\small{The grand potential for the four branches of black holes with spin-three chemical potential $\mu$. Here $\mu=1$, $T$ the temperature, and the grand potential is plotted as a function of the dimensionless parameter $\mu T$ .
} }\label{fvt}
\end{figure}

It was found in \cite{paper1} that  thermal boundary conditions for the ${\mathfrak {sl}}(3)$ Chern-Simons connections with spin three chemical potential admit {\em four} distinct branches, only one of which is smoothly connected to the BTZ black hole in the limit of vanishing chemical potential. The free energy of the BTZ branch can be expanded in powers of $\mu T$ and can be shown to agree precisely with results from large-$c$ limit of CFT  with ${W}_3$ symmetry. As displayed in figure \ref{fvt}, the BTZ branch merges with Branch II and then ceases to exist beyond a critical value of $(\mu T)$. Of the two remaining branches,  Branch III is thermodynamically favoured and its high $T$ limit expected to match the thermodynamics of a large-$c$ CFT with ${W}_3^{(2)}$ symmetry.  Our original motivation was to understand from a CFT perspective, certain intriguing aspects of this (UV) theory.

\subsection{$W^{(2)}_3$  algebra and semiclassical limit}
In this section we discuss relevant features of the Polyakov-Bershadsky algebra and its semiclassical limit. The main purpose of this section is to point out that the semiclassical limit of this algebra, in the Ramond sector, has an immediate interpretation which nontrivially matches the dual AdS bulk black hole description in terms of the flat connections of ${\rm SL}(3,{\mathbb R})\times {\rm SL}(3,{\mathbb R})$ Chern-Simons theory.

The best understood series of $W$-algebra CFTs are these with conventional $W_N$-algebras which can be seen to arise via the GKO coset construction \cite{GKO}, or  alternatively using Drinfel'd-Sokolov (DS) reduction \cite{DS}. Such reductions are assocated to ${\mathfrak {sl}}(2)$ embeddings in ${\mathfrak {sl}}(N)$ and conventional $W_N$-algebras are associated to the principal emebdding of ${\mathfrak {sl}}(2)$ in ${\mathfrak {sl}}(N)$.  However, the DS construction can be  generalized to other ${\mathfrak {sl}}(2)$ embeddings. For $N=3$, one such non-principal embedding yields the $W_3^{(2)}$-algebra. In a CFT with $W^{(2)}_3 \times W^{(2)}_3$ symmetry, the algebra is realized via the OPE of the  stress tensor $T(z)$, two spin-$\frac{3}{2}$ currents  $G^\pm(z)$ and a $U(1)$ current $J(z)$:
\begingroup
\allowdisplaybreaks
 \bea
 &&T(z)T(0)\,=\,\frac{c}{2z^4}+\frac{2T(0)}{z^2}+\frac{\partial T(0)}{z}+\cdots\ \label{w23ope}\\\nonumber\\\nonumber
&&T(z)J(0)\,=\,\frac{J(0)}{z^2}+\frac{\partial J(0)}z+\cdots\,, \qquad T(z)G^\pm(0)=\frac{3G^\pm(0)}{2z^2}+\frac{\partial G^\pm(0)}z+\cdots\,,
 \\\nonumber\\\nonumber
&&J(z)J(0)\,=\,\frac{\kappa}{z^2}+\cdots\ ,\qquad\qquad  \qquad\,\, J(z)G^\pm(0)\,=\,\pm\frac{G^\pm(0)}{z}+\cdots\,, \\\nonumber\\\nonumber
&&G^+(z)G^-(0)\,=\,\frac{( k-1)(2k -3)}{\,z^3}\,-\,\frac{3( k-1)J(0)}{\,z^2}\\\nonumber &&
\qquad\qquad\qquad\qquad\qquad\qquad+\,\frac{1}{z}\left((k-3){T(0)\,+\,3:J(0)^2:\,-\,\frac{3}{2}(k-3)\partial J(0)}\right)\,+\,\cdots\
 \eea
 \endgroup
 where $k$ is related to the level of the ${\mathfrak{sl}}(3)$ affine algebra and the central charge $c$ is given by\footnote{Usually, the $W_3^{(2)}$ algebra is written in terms of the affine algebra level $\hat k\,=\,-k$ (see e.g. \cite{Wyllard:2010rp})}
 \be
 c\,=\,\frac{(2k-3)(3k-1)}{k-3}\ ,\qquad \kappa\,=\,-\frac{2k-3}3\,.\label{chat}
 \ee
 In the context of AdS$_3$/CFT$_2$ duality for higher spin theory, $k$ is the level of the ${\rm SL}(3,{\mathbb R})\times {\rm SL}(3,{\mathbb R})$ Chern-Simons theory in the bulk dual. 
 The semiclassical limit for the latter requires 
 \be
 c\to \infty, \qquad k\to\infty\,,\qquad c\,=\,6 k\,.
 \ee
 In this limit however, the level of the $U(1)$ current algebra $\kappa\,\to\, -2k/3$ becomes negative and large and thus the corresponding CFT cannot be unitary. 
 
 Using the (Laurent) mode expansions for the currents, 
 \be
 J(z)\,=\,\sum_n z^{-n-1} J_n\,,\qquad G^\pm(z)\,=\,\sum_n z^{-n-\frac{3}{2}} G^\pm_n\,,\qquad T(z)\,=\,\sum_n z^{-n-2} L_n\,,
 \ee
we can represent the algebra in terms of the commutation relations for the modes \eqref{modexp}. Although the currents $G^\pm$ have half-integer spin, they are bosonic and the algebra is nonlinear\footnote{In this respect it differs crucially from the ${\cal N}=2$ superconformal algebra with which it bears some obvious similarity.}. The half-integer spin fields allow for both integer (Ramond) and half-integer (Neveu-Schwarz) modings.

\subsection{Ramond sector zero mode algebra and the semiclassical limit}
Let us focus attention on the Ramond sector where $G_n^\pm$ are integer moded, and in particular, on the zero mode sector spanned by $\{L_0, G_0^\pm, J_0\}$. The algebra of the zero modes,
\bea
\left[L_0, J_0\right]\,=\, \left[L_0, G_0^\pm\right]\,=\,0\,,\qquad \left[J_0, G_0^\pm\right]\,=\,\pm G_0^\pm\,,\\\nonumber\\\nonumber
\left[G_0^+, G_0^-\right]\,=\,-\frac{(k-1)(2k-3)}{8}\,+(k-3)L_0\,+\,3 J_0^2\,,
\eea
 has two Casimirs which label a given representation: the first is $L_0\,=\,\Delta\mathds{1}$ which yields the conformal dimension, while the second Casimir is
 \be
 \tfrac{1}{2}\{G_0^+, G_0^-\}\,+\,J_0^3\,+\,\left((k-3)L_0\, +\,\tfrac{1}{2}-\tfrac{(k-1)(2k-3)}{8}\right)\,J_0\,=\,\Delta^{(3)}\mathds{1}\,,
 \ee
 where the first term on the left hand side is the anticommutator of the two zero modes. The representations of the zero mode algebra have a structure that is quite similar to that of $SU(2)$. Highest weight states $|\psi\rangle$ in the Ramond sector come in degenerate multiplets spanned by the states $\left(G_0^+\right)^j|\psi\rangle$obtained by the action of the raising operator. For an irreducible representation ${\cal R}$ of dimension $n$, the generators of the zero mode algebra can be represented as $n\times n$ matrices (see \cite{Arakawa:2010sc}):
 \bea
 &&J_0\,=\,y\,{\mathds {1}}_{n\times n}\,+\,{\rm diag}\left(\tfrac{n-1}{2}, \,\tfrac{n-3}{2},\ldots,  -\tfrac{n-1}{2}\right)\,,\qquad L_0\,=\,\Delta\,\mathds{1}_{n\times n}
 \label{repnxn}\\\nonumber\\\nonumber
 && \left(G_0^+\right)_{\ell, j}\,=\,\sqrt{j(n-j)\left(3y+\tfrac{n}{2}-j\right)}\,\delta_{j,\ell+1}\,,\qquad G_0^-\,=\,\left(G_0^+\right)^\dagger\,,\qquad \ell=1,2 \ldots n-1\,.
  \eea
 The conformal dimension $\Delta$ and the second Casimir $\Delta^{(3)}$ can be epxressed in terms of the three quantum numbers $(k,n,y)$ as
 \be
 \Delta\,=\,\frac{1}{8(k-3)}\left((2k-3)(k-1)+2-2n^2-24y^2\right)\,,\qquad \Delta^{(3)}\,=\,\frac{1}{2}y\left(n^2-4y^2\right)\,.
 \ee
 In the large-$k$ semiclassical limit, the central charge $c \to 6k$, and letting both $n$ and $y$ scale with $k$, we find
 \be
\left.\Delta\right|_{k\gg 1}\to\,\frac{c}{24}\,-\,\frac{3n^2}{2c}\,-\,\frac{18y^2}{c}\,.
 \ee
 Thus the conformal weights of the  highest weight states are negative definite in this limit.  
 
 It is useful to rewrite the Casimirs in terms of the highest weight vectors ${\bf\Lambda}$ used to label the corresponding representations. In terms of these the conformal dimensions of the highest weight states (in the Ramond sector) is given  by \cite{Arakawa:2010sc}
 \be
 \Delta\,=\, \frac{{\bf \Lambda}\cdot\left({\bf\Lambda}\,+\,2 {\bm{\rho}}\right)}{2(3-k)}\,-\,\tilde{\bm\nu}\cdot{\bf\Lambda}\,+\,\frac{\kappa}{8}\,.
\ee
 Here ${\bm \rho}$ is the Weyl vector, given by the sum of the fundamental weights $({\bm w_1},{\bm w_2})$ and $\tilde {\bm \nu}={\bm w_2}$ for for ${\mathfrak{sl}}(3)$.  A convenient basis for the weight space of ${\mathfrak{sl}}(3)$ is provided by the three linearly dependent vectors $({\bm e_1},{\bm e_2},{\bm e_3})$ satisfying $\sum_i {\bm e_i}=0$ and\footnote{Explicit representation for ${\bm e}_i$ can be obtained by considering unit vectors $(\hat {\bm e}_1, \hat {\bm e}_2,\hat {\bm e}_3)$ on ${\mathbb R}^3$ and define ${\bm e_i} \,\equiv\,\hat{\bm e}_i\,-\,\hat{\bm \gamma}/3$ where $\hat {\bm \gamma}\,=\,\sum_i \hat{\bm e}_i$.} 
 \be
 \langle {\bm e_i}, {\bm e_j}\rangle\,=\,\delta_{ij}\,-\,\frac{1}{3}\,.
 \ee
 In terms of these, the fundamental weights of ${\mathfrak{sl}}(3)$ are ${\bm w}_1\,=\,{\bm e}_1$ and ${\bm w}_2\,=\,{\bm e}_1+{\bm e}_2$.
 The highest weight vector ${\bm \Lambda}$ can be expressed in this basis as
 \be
 {\bm\Lambda}\,=\,\left(-k+1+\frac{n}{2}+3y\right)\,{\bm e}_1\,+\,\left(-k+2-\frac{n}{2}+3y\right)\,{\bm e}_2\,.
 \ee
 From this, it is clear that the dimension of the representation is also determined compactly in terms of  ${\bm \Lambda}$,
 \be
 n\,=\,({\bm e}_1-{\bm e}_2)\cdot{\bm \Lambda}\,+\,1\,.
 \ee
 In the semiclassical limit of large $k$, it is thus natural to take both the dimension of the representation $n$ and the parameter $y$ to scale with $n$, so that 
 \be
{\bm \Lambda}\xrightarrow{k\to\infty} k\,{\bm\Lambda}_-\,,\qquad
{\bm\Lambda}_-\,\equiv\,\frac{n}{2k}({\bm e}_1-{\bm e}_2)-\left(\tfrac{3y}{k}-1\right)\,{\bm e}_3\,.
 \ee
 The Weyl vector for ${\mathfrak {sl}}(3)$ is independent of $k$ and thus can be neglected in the formula for $\Delta$ in this limit. An immediate consequence of this is that both the Casimirs $\Delta$ and $\Delta^{(3)}$ depend only on the combination ${\bm \omega}\,=\,{\bm \Lambda}_-+\tilde{\bm \nu}$. Viewing ${\bm \omega}$ as a vector in ${\mathbb R}^3$ with components $(\omega_1, \omega_2, \omega_3)$, we define the quantities
 \be
 C_s({\bm \omega})\,=\,\frac{1}{s}\sum_{i=1}^3\omega_i^s\,.
\ee
 Thus we may write the Casimir conditions for the ground states (in the semiclassical limit) as 
 \bea
 &&\langle L_0\rangle -\,\frac{c}{24}\,\xrightarrow{k\to\infty} \,-k\,C_2({\bm \omega})\,,\label{casimir}\\\nonumber\\\nonumber
 &&\Delta^{(3)}\mathds{1} \xrightarrow{k\to\infty}\left\langle\tfrac{1}{2}\{G_0^+, G_0^-\}\,+\,J_0^3\,+\,\frac{c}{6}\,J_0\left(L_0\,-\tfrac{c}{24}\right)\right\rangle\,=\,k^3\,C_3({\bm \omega})\,.
 \eea
 We note that in the semiclassical limit, the dimension of the representation becomes infinite as $n\sim {\cal O}(k)$, and the zero mode matrices $(J_0, G_0^\pm)$ effectively commute. In particular, the scaling of the off-diagional entries  in eq.\eqref{repnxn} implies that $G_0^\pm \sim {\cal O}(k^{3/2})$, while $L_0\sim {\cal O}(k)$ and $J_0 \sim {\cal O}(k)$ in the large-$k$ limit, and the commutator $[G_0^+, G_0^-]$ is thus subleading.  Therefore, the limit of large $k$  is semiclassical in nature, wherein the correspondence principle applies and the zero mode operators can be replaced by $c$-numbers and the Casimir conditions viewed as algebraic equations.
 
 We are interested in comparing the large-$k$ formulae with corresponding results for the spin-$\frac{3}{2}$ black hole obtained in \cite{paper1} using the Chern-Simons formulation.  In order to make this comparison, we perform certain rescalings and define
 \be
 \langle{G^\pm_0}\rangle\,=\,\pm i^{3/2}\pi\sqrt{\frac{k}{{2}}} {\cal G}^\pm\,,\qquad \langle J_0\rangle\,=\,iq,
 \qquad {\cal E}\,=\,\langle L_0\rangle\,-\,\frac{c}{24}\,.\label{zeromodedef}
 \ee
 The factors of $i$ appear when connecting the expectation values of the zero modes on the $z$-plane, to those on the cylinder using the map $z=e^{iw}$.  
 The minus sign and the factor of $\sqrt{k}$ have also been introduced in order to match with the conventions for the expectation values of the $W_3^{(2)}$ currents on the cylinder in \cite{paper1}.
 The Casimir conditions are thus,
 \bea
&& \boxed{{\cal E}\,=\,-k^{-1}\left(\frac{n^2}{4}\,+\,3 y^2\right)\,,}\label{casimirs}\\\nonumber\\\nonumber
 &&\boxed{\frac{\pi^2}{2}{\cal G}^+{\cal G}^-\,-\,k^{-1}q^3\,+\, q\,{\cal E}\,=\,-\frac{i}{2k}y(n^2-4y^2)\,.}
 \eea
 We will demonstrate two important results below:
 \begin{enumerate}
 \item{The left hand sides of the two Casimir equations are to be identified with the traces of the square and the cube of the  ${\rm SL}(3, \mathbb R)$ Chern-Simons connection which determines the higher spin dual gravity background.
 }
 \item{The flatness requirement for the Chern-Simons connection corresponds to picking the saddle point of the path integral over the manifold (at large $k$) of zero mode ground states in the Ramond sector.}
 \end{enumerate}
 \subsection{Chern-Simons bulk dual at large $c$}
The holographic description of a $W_3^{(2)}$ CFT at large central charge is given by Chern Simons theory with ${\rm SL}(3,{\mathbb R})\times {\rm SL}(3,{\mathbb R})$ gauge group and gauge connections $A(w, \bar w,\rho)$ and $\bar A(w,\bar w, \rho)$. The $\rho$ coordinate has the intepretation of the radial direction in ${\rm AdS}_3$ spacetime and for translationally invariant states in the CFT, the dependence can be factored out, and the connections replaced by gauge equivalent constant matrices $(a, \bar a)$:
\begin{align}
A\,=\, b^{-1} a \, b + b^{-1} db, \qquad \bar A \,=\, b \, \bar a \, b^{-1} + b \, db^{-1}\,, \qquad b \, =\, e^{\rho \hat{L}_0} \,.
\end{align}
Here $\hat L_0$ is the dilatation generator associated to the ${\mathfrak{sl}}(2)$ algebra in the appropriate non-principal embedding in ${\mathfrak{sl}}(3, {\mathbb R})$. The constant connections $(a, \bar a)$ each have two components:
\bea
a\,=\,a_w\,dw\,+\,a_{\bar w}\, d\bar w\,,\qquad \qquad\bar a\,=\,\bar a_w\,dw\,+\,\bar a_{\bar w} \,d\bar w\,,
\eea
where $a_{\bar w}$ and $\bar a_w$ are non-vanishing only when  a chemical potential $\lambda$ for spin-$\tfrac{3}{2}$  charge is turned on.
We define the matrices $\{\hat L_0, \hat L_{\pm 1}, J_0, G^\pm_{1/2}, G^\pm_{-1/2}\}$ in terms of ${\rm SL}(3, \mathbb R)$ generators as in \eqref{generators}. For the moment we ignore the chemical potential and focus attention on the holomorphic component $a_w$ and its antiholomorphic counterpart in the barred sector $\bar a_{\bar w}$. Following the conventions in \cite{paper1}, the relevant gauge connection (in highest weight gauge \cite{Gutperle:2011kf}) is:
\bea
a_w\,=\,\left[ \hat{L}_1 + \frac{3q}{2k}J_0 - \frac{w_{-2}}{k}\hat{L}_{-1} +\frac{\pi  }{\sqrt{2}k} \left( {\cal G}^+G^+_{-1/2}\,+\,{\cal G^-}G^-_{-1/2} \right) \right]\label{a}
\eea
and similarly in the barred sector
\bea
\bar a_{\bar w}\,=\,- \,\left[ \hat{L}_{-1} -\frac{3q}{2k} J_0 \,-\,\frac{w_{-2}}{k} \hat{L}_1 -\frac{\pi}{\sqrt{2}k} \left( {\cal G}^-G^-_{1/2}\,- \,{\cal G}^+G^+_{1/2} \right) \right]\,.
\label{bara}
\eea
In \cite{paper1}, by matching the Chern-Simons equations of motion to Ward identities of the $W_{3}^{(2)}$ algebra, it was shown that ${\cal G}^\pm$ and $q$ are the expectation values of the spin-$\frac{3}{2}$ currents and the $U(1)$ current respectively, while $w_{-2}$ is the expectation value of the `improved' stress tensor,
\be
w_{-2}\,=\left\langle-T(w)-\frac{3}{4k}J(w)^2\right\rangle\,=\,{\cal E}\,-\,\frac{3q^2}{4k} \,.
%\,=\,-{\cal E}\,-\,\frac{3}{4k}q^2\,.
\label{improved}
\ee 
Both $a_w$ and ${\bar a}_{\bar w}$ are  traceless. However, the higher traces are non-vanishing:
\be\label{holovacua}
\boxed{\tfrac{1}{2}{\rm Tr} \,a_w^2\,=\,\tfrac{1}k\left(w_{-2}\,+\,\tfrac{3q^2}{4k}\right) }\,,\qquad 
\boxed{\tfrac{1}{3}{\rm Tr}\, a_w^3\,=\,\tfrac1{k^2}\left(\tfrac{\pi^2}{2}\,{\cal G}^+{\cal G}^-\,-\,\tfrac{q^3}{4k}\,+\,{w_{-2}}\,q\right)}\,.
\ee 
It is clear that these have the same form as the quadratic and cubic Casimirs \eqref{casimirs} of the zero mode algebra of $W_3^{(2)}$ we encountered above, upon making the identification,
\be
{\cal E}\,=\,\left(w_{-2}\,+\,\tfrac{3q^2}{4k}\right) \,,
\ee
exactly as indicated by eq.\eqref{improved}.

  The Chern-Simons bulk description of the CFT state requires two further inputs. The first is the flatness of the connection, and the second is the smoothness requirement which constrains the holonomy of the gauge fields around the compact spatial direction or the compact (Euclidean) temporal circle in the thermal state.  When the holonomy is nontrivial (or non-smooth) we interpret the corresponding state as a conical deficit/excess. The flatness condition is simple in the absence of a chemical potential (i.e. when $a_{\bar w}=\bar a_w=0$). 
At zero temperature, the holonomy of the gauge field around the spatial circle must be trivial, corresponding to the higher spin version of global ${\rm AdS}_3$,
\be
{\rm Hol}_{\phi}(A)\,\equiv\, {\cal P}\exp\left(\oint A_\phi\right)\,=\,\mathds{1}\,,
\ee
where 
\be
A_\phi\,=\,A_w\,-\,A_{\bar w}\,,\qquad  A_t\,=\,i\left(A_w\,+\,A_{\bar w}\right)\,.
\ee
For the zero temperature ground state, this fixes the eigenvalues of $a_\phi$:
\be
{\rm spec}(a_\phi)\,=\,(i,0,-i)\,.
\ee
Since $a_{\bar w}=0$ in the absence of chemical potential deformatons, this implies that the ground state corresponds to an operator $\Phi$ with non-trivial (negative) conformal dimension
\be
{\cal E}\,=\, - \frac{c}{6}\,\implies\,\Delta_\Phi \,=\, \frac{c}{24}\,-\,\frac{c}{6}\,=\,-\frac{c}{8}\,.\label{negativew23}
\ee
Note that there are two zero mode states labelled by the quantum numbers $(y, n)$ with this energy and holonomy in the semiclassical limit, namely $(y, n)\,=\,(0, 2k)$ and $(y, n)\,=\,(k/2, \, k)$.  Both allowed solutions are highly degenerate and in the large $k$ limit, the family of ground states is  characterized by the algebraic variety \eqref{casimirs}.

Next we turn to the CFT in the thermal state. At high temperatures in the semiclassical limit, modular invariance tells us that the spatial circle at zero temperature is swapped for the Euclidean thermal circle with circumference $\beta$.  However, since the zero temperature ground state(s) have nontrivial conformal dimension $\Delta_{\Phi}\,=\, -\frac{c}{8}$ \eqref{negativew23}, we expect that the thermal cylinder is accompanied by insertions of a primary operator $\Phi$ at spatial infinity. Furthermore the manifold of ground states at zero temperature (given by the algebraic variety at large-$c$ spanned by the zero mode sector) now has a thermal counterpart given by the smoothness conditions for the holonomy around the thermal circle:
\be
{\rm Hol}_t(A)\,=\,{\cal P}\exp\left(\oint A_t\right)\,=\,\mathds{1}\, \implies \,{\rm spec}(a_t)\,=\,\left( 2\pi i\beta^{-1},\,0,\, -2\pi i\beta^{-1}\right)\,.\label{spectherm}
\ee
Therefore the constraints on the quadratic and cubic Casimirs are\footnote{Note that $a_w \,=\,-i a_t$ when $\lambda=0$.},
\be
{\cal E}\,=\,k\frac{4\pi^2}{\beta^2}\,\qquad\qquad \frac{\pi^2}{2}{\cal G}^+{\cal G}^-\,-\,k^{-1}q^3\,+\,{\cal E}q\,=\,0\,.\label{hightconstraints}
\ee
With $k\,=\, c /6$, this is the expected high temperature behaviour for the CFT with rescaled temperature $\beta_\Phi\,=\,\beta/2$ as in \eqref{enden}.
\subsection{Grand canonical saddle point and flat C-S connections}
We now introduce a bias in the system via a small chemical potential for the spin-$\tfrac{3}{2}$ currents. At the level of the CFT action this is a holomorphic (plus antiholomorphic) deformation
\be
\delta I_{\rm CFT}\,=\,-\lambda \int_{S^1\times{\mathbb R}_t} d^2w\,\left(g_+\,{G^+}(w)\,+\,
g_-\,{G^-}(w)\right)\,+\,{ \rm h.c.}\label{chemical}
\ee
where $\lambda$ is a dimension-$\frac{1}{2}$ chemical potential and $g_\pm$ are dimensionless parameters.   For simplicity we set $g_+ = g_-$.  In the dual bulk Chern-Simons theory, the deformation modifies the gauge connections so that $a_{\bar w}\neq 0$ and ${\bar a}_{ w}\neq 0$ as in \cite{paper1},
%When a common chemical potential $\lambda\neq0$ is introduced for the two spin-$\frac{3}{2}$ charges, 
%the connections are modified with 
\bea
%a&& = \, \left[ \hat{L}_1 + \frac{3q}{2k}J_0 - \frac{w_{-2}}{k}\hat{L}_{-1} +\frac{\pi  {\cal G}}{\sqrt{2}k} \left( G^+_{-1/2}+G^-_{-1/2} \right) \right] dz\\\nonumber\\\nonumber
&& a_{\bar w}\,=\, \,\lambda \left[ \sqrt{2} \left( G^-_{1/2} - G^+_{1/2} \right) + \frac{3 q}{\sqrt{2}k} \left( G^+_{-1/2}+G^-_{-1/2} \right) +\frac{\pi {\cal G}}{k}\hat{L}_{-1} \right]  \label{w23_connection}\\\nonumber\\\nonumber
&&\bar{a}_w\, =\,  \lambda \left[ \sqrt{2} \left( G^+_{-1/2}+G^-_{-1/2} \right) + \frac{3q}{\sqrt{2}k} \left( G^-_{1/2} - G^+_{1/2} \right) -\frac{\pi{\cal G}}{k}\hat{L}_1  \right]\,,
\eea
and the flatness condition $[a_w, a_{\bar w}]\,=\,[\bar a_w,{\bar a}_{\bar w}]\,=\, 0$, then constrains the expectation values of the improved stress tensor \eqref{improved} and the spin-$\tfrac{3}{2}$ charges,
\be
w_{-2}\,=\,\frac{ 9 q^2}{4k}\,,\qquad\qquad {\cal G}\,=\,\langle g_+G^+\rangle\,=\,\langle g_- G^-\rangle\,.\label{flatcond}
\ee
Interestingly, both  conditions can be understood as a consequence of a large-$k$ saddle point of the path integral over the zero mode ground states in the Ramond sector. In particular, let us  consider  schematically the path integral (at large $k$) over the zero modes. The small $\lambda$ limit is best understood as a high temperature limit where $\lambda\sqrt\beta \ll 1$ and we make use of the high temperature versions of the Casimir conditions (or holonomies)
\bea
&&{\cal Z}\,\sim\,\int d{\cal E}\, dq\, d{\cal G}^+\, d{\cal G}^-\,\times\\\nonumber\\\nonumber
&&\delta\left({\cal E}\,-\, \tfrac{4\pi^2}{\beta^2} k\right)\,\delta\left(\frac{\pi^2}{2}{\cal G}^+{\cal G}^- - k^{-1}q^3 + q{\cal E}\,\right)\, e^{-\beta\ell\lambda\left ({\cal G}^+ +{\cal G}^-\right)}\,\ldots
\eea
The ellipsis represent any other factors in the partition sum including putative higher order terms in $\lambda$ and we have introduced an explicit dimensionful factor $\ell$ representing the size of the spatial circle. Also omitted are finite $\lambda$ corrections to the holonomy conditions which  can be found in \cite{paper1}.
Implementing the delta-functions constraints via Lagrange multipliers and locating  the saddle point   with respect to $q$, ${\cal G^\pm}$, ${\cal E}$ and the Lagrange multipliers, which dominates the integral at large $k$ and high temperatures, we find in the $\lambda\to 0$ limit:
\be
{\cal E}\,=\, \frac{3}{k} q^2\,,\qquad {\cal G}^+\,=\,{\cal G}^-\,,
\ee
which, after using \eqref{improved}, reduces precisely to the conditions \eqref{flatcond} resulting from imposing the flatness of the Chern-Simons connection at finite (small) chemical potential $\lambda$\footnote{We note that the saddle point or C-S flatness condition leads to the constraint $w_{-2}= 9q^2/4k$, which is incompatible for {\em real} non-zero $q$ with the zero temperature holonomy condition \eqref{holovacua} which requires $w_{-2}<0$.  }.

To summarize, we have provided evidence above  that the classical, large-$c$ holographic dual of a  CFT with ${W}_3^{(2)}$ symmetry has a large degeneracy of states which appear as a continuum in the large-$c$ limit, with a large negative conformal dimension. The large-$c$ manifold of ground states is given by the algebraic variety \eqref{casimirs}. 

\subsection{Ward identities for finite $\lambda$}
In this subsection, for the sake of completeness, we recall the analysis presented in \cite{paper1} where it was shown that  Ward identities of the large-$c$ ${ W}^{(2)}_3$ CFT with spin-$\tfrac32$ chemical potential $\lambda$, are reproduced by the bulk Chern-Simons field equations,
\begin{align}
da+a \wedge a =0, \qquad d\bar{a} + \bar{a} \wedge \bar{a} =0.
\end{align}
We impose the flatness conditions on the connections $a= a_w\, dw \,+\, a_{\bar w}\, d\bar w$ and $\bar a=\bar a_w\, dw \,+\, \bar a_{\bar w}\, d\bar w$ treating their parameters $(q, w_{-2}, {\cal G})$
as functions of $(w,\bar w)$ with $w\,=\,\phi\,+\,it$. This yields the set of first order equations:
\begin{eqnarray}\label{eq_bulk}
&&\partial_{\bar{w}} \left( w_{-2}\,+\,\frac{3}{4k}\,q^2 \right) \,=\, -\pi \lambda\, \partial_z {\cal G}, \qquad\qquad \partial_{\bar{z}} q = -4k\lambda \,w_{-1} \,,\\\nonumber
&&\partial_{\bar{w}} w_{-1} \,=\, \tfrac\lambda k \left( -w_{-2} +\frac{9}{4k}q^2 \right) ,\qquad\qquad 
\partial_{\bar{w}} {\cal G} \,=\, \tfrac{3}{\pi}\, \lambda \,\partial_w q\,. \nonumber
\eea
These conditions are equivalent to the Ward identities of the CFT with ${ W}^{(2)}_3$ symmetry, deformed by a spin-$3/2$ current,
\begin{align} \label{deformation}
\delta I_{\rm CFT}\, =\, - \int_{S^1\times{\mathbb R}_t} d^2w \lambda(\bar{w}) \left(g_+ G^+(w) + g_- G^-(w) \right)\,.
\end{align}
 The Ward identities of this deformed theory are obtained by computing the expectation values of the left hand sides of \eqref{eq_bulk} pertubatively in $\lambda$, using the identity $\partial_{\bar{w}} \left( \frac{1}{w}\right) \,=\, 2\pi \delta^2(w,\bar{w})$ and the OPEs of the currents \eqref{w23ope},
\begin{align} \label{ward}
&\tfrac{1}{2\pi} \partial_{\bar{w}} \langle T(w) \rangle_\lambda\,  =\, \tfrac{1}{2} \lambda \, \partial_w \langle g_+ G^+ (w)\, +\, g_- G^- (w) \rangle
\\\nonumber
\\
&\tfrac{1}{2\pi} \partial_{\bar{w}} \langle J(w) \rangle_\lambda  \,=\, - \lambda \, \langle g_+ G^+ (w)\, -\, g_- G^- (w) \rangle \nonumber
\\\nonumber
\\
&\tfrac{1}{2\pi} \partial_{\bar{w}} \langle g_+ G^+ (w) \,- \,g_- G^- (w) \rangle_\lambda \, =\, 2 g_+ g_- \lambda \, \langle T(w) \,+\, \tfrac{18}{c} J(w)^2 \rangle\nonumber
\\\nonumber
\\
&\tfrac{1}{2\pi} \partial_{\bar{w}} \langle g_+ G^+ (w) + g_- G^- (w) \rangle_\lambda \, =\, 3 g_+ g_- \lambda \, \partial_w\langle J(w)\rangle.\nonumber
\end{align}
These Ward identities show that the $W_3^{(2)}$ currents are not actually  holomorphic after the  chiral deformation  \eqref{deformation} is introduced.
Comparing these with the bulk field equations \eqref{eq_bulk}, the CFT expectation values of the currents can be identified in terms of bulk quantities,
\begin{align} \label{identifications}
&w_{-1} \,=\, \frac{\pi}{2k} \langle g_+ G^+ (w)\, -\, g_- G^- (w) \rangle_\lambda, \qquad {\cal G}\,=\,\langle g_+ G^+ (w) + g_- G^- (w) \rangle_\lambda\\
&w_{-2} = \langle -T(w) \,-\,\frac{3}{4k} J(w)^2 \rangle_\lambda, \qquad\qquad q\, = \,\langle J(w) \rangle_\lambda, \qquad g_+g_- = \frac{1}{2\pi^2}.\nonumber
\end{align}
In the following sections, we will consider situations both with $\lambda$ zero and non-zero from within the holographic large-$c$ description, and match with our CFT interpretation.
% the theory in both the c\eqref{w23_connection} for $\lambda=0$. Therefore, the subscript in the expectation values, $\langle \dots \rangle_{\lambda =0}$, will be suppressed, $\langle \dots \rangle$.

\section{Zero charge $W^{(2)}_3$ black hole}
\subsection{Thermodynamics}
We now  turn our attention to the thermodynamical properties of the holographic large-$c$ theory in the canonical ensemble ($\lambda=0$).   We have seen that the high temperature energy of this thermal state is given by (summing over both holomorphic and antiholomorphic sectors) 
\be
-\langle T \rangle\,-\,\langle \bar T \rangle\,=\,2{\cal E}\,=\,c\,\frac{4\pi^2}{3\beta^2}\,, 
\ee
 which then corresponds to the thermal entropy 
 \be
 S\,=\,c\,\frac{8\pi^2}{3\beta}\,.\label{s_thermal}
 \ee
 Both quantities above are a factor of 4 times the expected result for a BTZ black hole (or high temperature CFT)  at temperature $\beta^{-1}$. The constant Chern-Simons connections for the zero charge black hole in the non-principal embedding are:
 \begin{equation}
\label{w23_conn0}
a  = \left(\hat{L}_1 - \frac{w_{-2}}{k} \hat{L}_{-1} \right) dw, \qquad \bar{a}  = - \left( \hat{L}_{-1} - \frac{w_{-2}}{k} \hat{L}_1 \right) d\bar{w}\,.
\end{equation} 
 These have the same form as those for the BTZ black hole, the crucial difference being the non-principal embedding of ${\mathfrak{sl}}(2)$ generators $\{\hat L_{\pm 1}, \hat L_0\}$ in the ${\mathfrak{sl}}(3)$ algebra (see appendix \ref{generators}).  Our interpretation of this difference is that it is a consequence of the large (negative) conformal dimension of the zero temperature ground state.  The high temperature expectation value of the stress tensor in this situation follows from the CFT result \eqref{onepthighT},
 \be
 \langle T \rangle_{\beta,{\Phi}}
 \,= \,\left( \frac{2\pi }{\beta}\right) ^2 \Delta_\Phi \,-\, \frac{c\pi^2}{ 6 \beta^2}\,=\,-c\,\frac{2\pi^2}{3\beta^2}\,,
 \ee
 where $\Delta_\Phi\,=\,-\frac{c}{8}$ is the conformal dimension\footnote{\label{twist2}The twist squared operator, $\sigma^2_n$, whose two point function measures entanglement negativity \cite{Calabrese:2012ew}, has the same conformal dimension as $\Phi$ in eq.\eqref{negativew23},
\begin{align}
\lim_{n_e \to 1} \Delta_{\sigma_{n_e}^2} =  - \frac{c}{8},
\end{align}
where $n_e$ is the even replica index. Note that this is the weight associated to one chiral sector. We thank V. Malvimat for drawing our attention to this.}of the zero temperature ground state \eqref{negativew23}. The insertion of this primary on the high temperature thermal  cylinder leads to the standard high temperature thermodynamics, but at a {\em larger}  rescaled temperature $\beta_\Phi\,=\, \beta/2$.

\subsection{Entanglement entropy}

\paragraph{From holography:} The prescription for computing the entanglement entropy in the Chern-Simons formulation of gravity with higher spin fields involves the calculation of a bulk (AdS$_3$) Wilson line in a specific representation ${\cal R}$, anchored at the endpoints $P$ and $Q$ of the entangling interval on the boundary CFT \cite{deBoer:2013vca, Ammon:2013hba},
\bea
&&S_{\rm EE}(P, Q)\,=\,\frac{k}{\sigma_{1/2}} \,\ln\left[\lim_{\rho_P, \rho_Q \to \infty} W_{\cal R}(P,Q)\right]\,,\\\nonumber\\\nonumber
&&W_{\cal R}(P, Q)\,=\,{\rm Tr}_{\cal R}\left[{\cal P}\,\exp\left(\int_P^Q \bar A \right)\exp\left(\int_Q^P A \right)\right]\,.
\eea
The proposal of \cite{deBoer:2013vca} included an overall normalization constant $\sigma_{1/2}=2$ when the embedding yields half-integer spin currents, otherwise $\sigma_{1/2}=1$. The constant was introduced in order that the entanglement entropy  agree with  the thermal entropy for large intervals when the former becomes extensive\footnote{In \cite{deBoer:2013vca} the connections corresponding to the zero charge black hole were also studied but with the holonomy condition  ${\rm spec}(a_t)\,=\,(i\pi\beta^{-1},0, -i\pi\beta^{-1})$. The corresponding holonomy is trivial in the truncation to  ${\rm SL}(2,{\mathbb R}) \times U(1)$. In order to obtain the identity element of ${\rm SL}(3,{\mathbb R})$ we need the condition \eqref{spectherm}. }. However, in the present context we will see below that this normalization is necessary for agreement with the short interval limit which is determined by the universal OPE for the twist operators \eqref{opetwi}.

Although this will not be important for us in this paper, the prescription for the Wilson line which matches the CFT evaluation of entanglement entropies  \cite{Datta:2014ska, Datta:2014uxa} is the so-called ``holomorphic" prescription \cite{deBoer:2013vca}. The other ``canonical" prescription reduces to the holomorphic one when $A_{\bar w}\,=\,\bar A_w\,=\,0$ which is the case when the chemical potential $\lambda$ is vanishing. In the ${\rm SL}(3,{\mathbb R})\times {\rm SL}(3,{\mathbb R})$ Chern-Simons theory the entanglement entropy for both the principal and non-principal embeddings is 
given by the Wilson line in the adjoint representation \cite{deBoer:2013vca},
\be
{\rm dim}[{\cal R}]\,=\,8\,.
\ee
The Wilson line  in the adjoint representation is obtained by the product of the Wilson lines in the fundamental and conjugate representations,
\bea
&&W_{\rm fund}\,=\,\lim_{\rho_P,\rho_Q \to \infty}\text{Tr} \left[ e^{\rho_Q \hat L_0}\, {\rm exp} \left(-\Delta\bar{w} \, \bar{a}\right) e^{-2\rho_P \hat L_0}\, {\rm exp} \left( \Delta w\, a \right) e^{\rho_Q \hat L_0} \right]\,,\\\nonumber\\\nonumber
&&W_{\rm Ad}\,=\,W_{\rm fund}\,W_{\overline {\rm fund}}\,.
\eea
The anti-fundamental Wilson line is identical to the fundamental Wilson line but with signs reversed in front of $\Delta w$ and $\Delta\bar w$. For the single interval entanglement entropy, $\Delta w=\Delta\bar w = L$, the size of the interval.  In the limit of large $\rho_{P,Q}$ wherein the endpoints of the Wilson line are anchored to the boundary of AdS$_3$, we obtain,
\be
W_{\rm Ad}\,=\,e^{2(\rho_P+\rho_Q)}\,\frac{k^2}{w_{-2}^2} \sinh^4 \left( \sqrt{\frac{w_{-2}}{k}}L \right)\,.
\ee
When the $U(1)$ charge is vanishing, the parameter $w_{-2}$ is identified with the energy of one chiral sector as in \eqref{improved}, and its value determined by the holonomy condition \eqref{hightconstraints} at high temperature. After subtracting out the additive divergent piece proportional to the UV cutoff, the EE for a single interval of length $L$  is,
\begin{align} \label{EE_hol}
S_{\rm EE}\,  =\, \frac{c}{3} \ln \left[\frac{\beta}{2\pi} \sinh \left( \frac{2\pi L}{\beta} \right) \right]\,.\qquad 
\end{align}
The standard formula for the high temperature EE for a CFT is $S_{\rm EE}\,=\,\frac{c}{3}\ln\left(\sinh\frac{\pi L}{\beta}\right)$. The result for the $W_3^{(2)}$ CFT  shows a subtle difference. The  argument of the hyperbolic sine function is twice the usual result.  We now interpret this as a consequence of the nontrivial ground state with $\Delta_\Phi\,=\,-\frac{c}{8}$ which manifests itself as a conical deficit or rescaled temperature  
$\beta_\Phi=\beta/2$ at high temperatures.
It is clear that the EE for a general conical deficit  \eqref{renyi}, does not yield  the thermal entropy \eqref{sphi} in the excited state in the limit of large interval size.
However, the limit of small interval size precisely matches the OPE of twist operators \eqref{opetwi}.

\paragraph{Entanglement entropy from CFT:}
We can evaluate the R\'enyi entropy in the CFT with ${\cal W}^{(2)}_3$ symmetry where the spin-$3/2$ and $U(1)$ charges are turned off.  This is a non-standard situation because the ground state of the theory has a large negative conformal dimension. The conformal dimension of the twist fields cannot immediately be inferred by  computing the stress tensor one-point function using the uniformisation map from a multi-sheeted Riemann surface implementing the replica trick. Each sheet is now accompanied by a pair of insertions of the primary field with conformal dimension $\Delta_\Phi$ corresponding to the nontrivial ground state. Instead we may view the R\'enyi entropy as a four-point correlator involving the two light branch point twist fields and two heavy primaries with $\Delta_\Phi\,=\,-\frac{c}{8}$.  Assuming only the vacuum block contribution to the HHLL correlator, the R\'enyi entropy in the large-$c$ limit can be obtained as in section \ref{sec:renyi} and in the $n\to 1$ limit,
\be
S_{\rm EE}\,=\,\frac{c}{3}\ln\left[\frac{\beta}{\pi \alpha_\Phi}\sinh\left(\alpha_\Phi \frac{\pi L}{\beta}\right)\right]\,,\qquad \alpha_\Phi\, =\,\sqrt{1-24\frac{\Delta_{\Phi}}{c}} \,.
\ee
Given $\Delta_\Phi=-\frac{c}{8}$, we recover the holographic EE \eqref{EE_hol}. 

\section{$W^{(2)}_3$ black hole with spin-$3/2$ and $U(1)$ charges}
The holonomy conditions \eqref{hightconstraints} at high temperature show that the thermal states at fixed energy are highly degenerate, characterized by condensates or expectation values for the spin-$\tfrac32$ and $U(1)$ charges. In general, we expect that the entanglement entropy will depend on the choice of state. However, the quantum theory sums over all these states and the resulting saddle point at large $k$ determines the thermodynamics.  It is nevertheless interesting and instructive to consider any given solution to the holonomy conditions and compute the holographic EE for that configuration.
\subsection{Holographic entanglement entropy for $W_3^{(2)}$ CFT}
The general result for the $W_3^{(2)}$ Wilson line in the fundamental representation  in terms of the 
eigenvalues of $a_w$ is,
\bea\label{WL_holographic_3/2}
{W}_{\rm fund}\, = &&\frac{e^{\rho_P+\rho_Q}}{(\lambda_1 \lambda_2 \lambda_3)^2}\times\\
 &&\left[ \lambda_1 \left( \frac{q}{k}+\nu_3 \right)e^{\nu_3 \Delta w} + \lambda_2 \left( \frac{q}{k}+\nu_1 \right)e^{\nu_1 \Delta w} + \lambda_3 \left( \frac{q}{k}+\nu_2 \right)e^{\nu_2 \Delta w}  \right] \times\nonumber \\
&&  \left[ \lambda_1 \left( \frac{q}{k}+\nu_3 \right)e^{-\nu_3 \Delta\bar w} + \lambda_2 \left( \frac{q}{k}+\nu_1 \right)e^{-\nu_1 \Delta\bar w} + \lambda_3 \left( \frac{q}{k}+\nu_2 \right)e^{-\nu_2 \Delta\bar w}  \right] \nonumber
\eea
where $\nu_{1,2,3}$ are eigenvalues of the matrix $a_w$ and the charges in the barred \eqref{a} and unbarred \eqref{bara} sectors are identical. For the single interval entropy, $\Delta w= \Delta\bar w= L$ and we have defined  
\begin{align}
\lambda_1 = \nu_1-\nu_2, \qquad \lambda_2 = \nu_2-\nu_3, \qquad \lambda_3 = \nu_3-\nu_1. 
\end{align}
Since the barred and unbarred sectors are identical, the Wilson line in the antifundamental representation yields the same expression and $W_{\rm Ad} \,=\,\left(W_{\rm fund}\right)^2$.
The expression  above was obtained without imposing the holonomy conditions and applies  for any values of $q, \, w_{-2}$ and ${\cal G}$. The $\{\nu_i\}$ are the roots of the cubic,
\be
\nu^3\,-\,\frac{1}{k}\nu\left(w_{-2}\,+\,\frac{3q^2}{4k}\right)\,-\,\frac{1}{k^2}\left({\cal G}^+{\cal G}^-\frac{\pi^2}{2}\,-\,\frac{q^3}{4k}\,+\,q w_{-2}\right)\,=\,0\,.
\ee
The holonomy conditions \eqref{hightconstraints} set the $\nu$-independent term in the cubic  to zero and fix  the coefficient of the term linear in $\nu$, so that the three eigenvalues are simply, 
\begin{align} \label{eval}
\left( \nu_1, \nu_2, \nu_3 \right) = \left( -\tfrac{2\pi}{\beta} , 0, \tfrac{2\pi}{\beta} \right)\, ,
\end{align}
for all values of $q, \, w_{-2}$ and ${\cal G}$ which satisfy the holonomy conditions.
The holographic entanglement entropy is then expressible in a compact form for generic $q$:
\be
S_{\rm EE}\,=\,\frac{c}{6}\ln\left[\frac{\beta^4}{(2\pi )^4}\sinh^4\left(\frac{2\pi L}{\beta}\right)\left(1-\frac{\beta^2}{4\pi^2}\frac{q^2}{k^2}{\rm tanh}^2\left(\frac{\pi L}{\beta}\right)\right)^2\right]\,.
\ee
We list the  salient features of this result:
\begin{itemize}  
\item{When $q=0$, it reduces to eq.\eqref{EE_hol} as expected. The zero temperature version of this, for the CFT on the unit circle is obtained by the replacement $\beta \to -2\pi i$.}
\item{The expression  implies an upper bound on the $U(1)$ charge $q$ in terms of the temperature.  Since $\tanh^2\tfrac{\pi L}{\beta}$ is monotonic in $L$, approaching unity as $L\to \infty$}, if  $|q|$ is too large the argument of the logarithm will have a zero at some finite value of $L$ and the EE will diverge at that point. In order to avoid this unphysical situation the
$U(1)$ charge must be bounded:
\be
|q| \leq \frac{2\pi}{\beta} k\,.
\ee
Interestingly, this bound is automatically implied by the dimension of the representation of the zero mode algebra in the semiclassical limit. Taking $y=0$ in eq.\eqref{repnxn} the magnitude of the largest eigenvalue of $J_0$ is $n/2$ (taking $n$ large). From \eqref{zeromodedef} and the Casimir condition \eqref{casimirs} it immediately follows that
\be
q_{\rm max}^2\,=\,k{\cal E}\,=\,\left(\frac{2\pi k}{\beta}\right)^2\,.
\ee

\item{For generic values of $q$ below this bound, for large enough $L$, the EE is extensive but is lower than the thermal entropy  \eqref{s_thermal} by a factor of two. This is a special case of what we have already noted for the EE of a conical deficit state (in the large-$c$ theory) \eqref{renyi}.}
\item{When the bound is saturated i.e. $|q| = 2\pi k/\beta$, the expression for EE simplifies and surprisingly reduces to the  standard CFT formula at finite temperature :
\be
S_{\rm EE}\,\big|_{|q| = \frac{2\pi k}{\beta}}\,=\,\frac{c}{3}\,\ln\left[\frac\beta\pi\,\sinh\left(\frac{\pi L}{\beta}\right)\right]\,.
\ee
Importantly, the thermal entropy for the canonical ensemble is still given by \eqref{s_thermal} and does not agree with the EE of this particular configuration in the large $L$ limit.
}
\item{The smoothness/holonomy conditions \eqref{hightconstraints} on the Chern-Simons connections relate the spin-$\tfrac{3}{2}$ charge to the $U(1)$ charge at a given temperature according to
\be
\frac{\pi^2}{2}{\cal G}^+{\cal G}^-\,=\, \frac{q}{k}\left(q^2\,-\,
\frac{4\pi^2}{\beta^2}k^2\right)\,.
\ee
\begin{figure}[h]
\begin{center}
\includegraphics[width=1.8in]{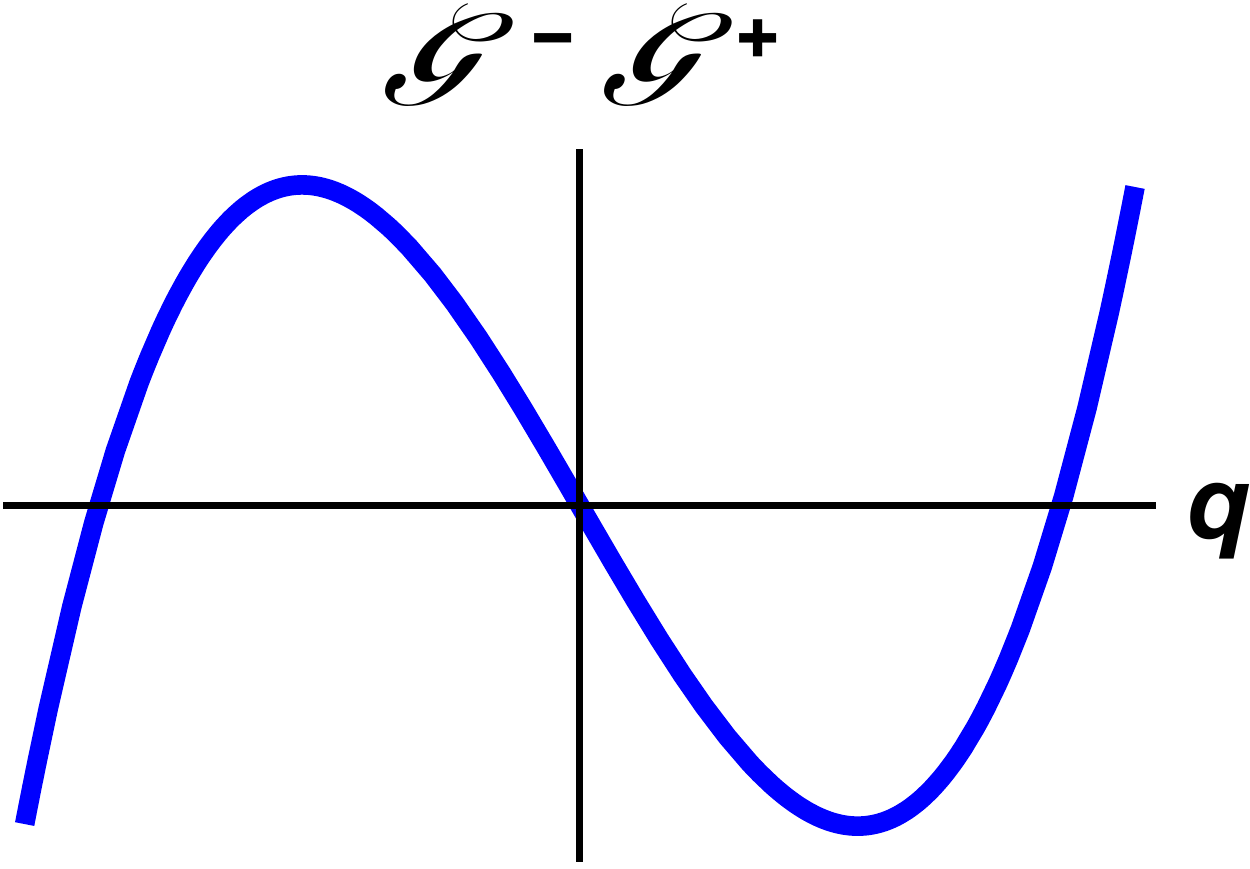}
\end{center}
\caption{\small{The product ${\cal G}^+{\cal G}^-$ of the spin-$\tfrac32$ charges versus the $U(1)$ charge $q$ for holographic backgrounds satisfying the holonomy conditions.  ${\cal G}^+{\cal G}^-$ has extrema when $q^2 = \tfrac43 \pi^2k^2\beta^{-2}$.
} }\label{gvq}
\end{figure}
Therefore the (product of) the spin-$\tfrac32$ charges increases from zero at $q=0$, reaches a maximum at $q^2= 4\pi^2k^2/3\beta^2$ and subsequently decreases to zero at $|q| = 2\pi k/\beta$ (see figure \ref{gvq}).
}
\item{It is interesting to note what happens when an arbitrarily small spin-$\tfrac{3}{2}$ chemical potential $\lambda$ is introduced. The holonomy constraints \eqref{hightconstraints} are then augmented with the Chern-Simons flatness (or large-$k$ saddle point) conditions \eqref{flatcond} and this fixes the value of the $U(1)$  charge precisely at $q^2 = \tfrac43 \pi^2k^2\beta^{-2}$, one of the two extrema of figure \ref{gvq}.}
\item{The expansion of the holographic EE in the limit of small interval lengths has  a nice interpretation:
\bea
S_{\rm EE}\big|_{\frac{L}{\beta}\ll 1}\,=\,\frac{c}{3}\,\ln L\,+\,\frac{1}{3}\left(\frac{4\pi^2}{\beta^2}k\,-\,\frac{3 q^2}{4k}\right) L^2\,+\ldots
\eea
We immediately recognize the $O(L^2)$ term as the expectation value of a particular combination of the stress tensor and the $U(1)$ current in the thermal state (with insertions of the operator $\Phi$ at spatial infinity),
\be\label{hol_EE_short}
S_{\rm EE}\big|_{\frac{L}{\beta}\ll 1}\,=\,\frac{c}{3}\left(\ln \, L\,-\, \frac{L^2}{c}\left\langle T(w)\,+\,\frac {3}{4k} J(w)^2\right\rangle_\Phi\,+\ldots\right)\,.
\ee
This has a natural interpretation in the CFT language as we now discuss.
 }
\end{itemize}

\subsection{Entanglement entropy from CFT at small $L$}
Let us revisit  the four-point correlator on the thermal cylinder which computes R\'enyi/entanglement entropy in CFT,
\begin{align}\label{renyi2}
{\cal C}_4\, =\, \frac{\langle \Phi(\infty) \, {\bar \sigma_n}(L) {\sigma_n}(0) \, \Phi(-\infty) \rangle}{\langle \Phi(\infty) \, \Phi(-\infty) \rangle}\,.
\end{align}
The presence of the $U(1)$ current in the $W_3^{(2)}$ algebra permits the appearance of an additional quasi-primary $J^2$ of dimension two in the twist-antitwist OPE. This means that on general grounds  the OPE of the twist fields must be of the form,
\begin{align}
 {\bar \sigma_n}(L) {\sigma_n}(0) \sim \frac{1}{L^{4h_{\sigma}}} \left( 1+ \frac{2h_{\sigma}}{c} L^2 \left( T(0)\,+\,a\,J^2(0)\,+{\bar T}(0)\,  + \,a\,{ \bar J}^2(0)\right)\ldots\right)\,,
\end{align}
where $a$ is a dimensionless constant.  There is a natural choice of $a$ from the viewpoint of the bulk Chern-Simons theory wherein the generators of the asymptotic symmetries of the metric commute with the generators of the $U(1)$ transformations.  This corresponds to the choice of an effective stress tensor $T_{\rm eff}$ which has a trivial OPE with $J(z)$,
\be
T_{\rm eff}(w)\,=\, T(w)\,+\,a\, J(w)^2\,,\qquad\qquad T_{\rm eff}(w) J(0)\,\sim\, {\rm regular}\,.
\ee
From the algebra \eqref{w23ope}, we can use the relevant large $k$ OPEs 
\be
T(w) J(0)\,\sim\,\frac{J(0)}{w^2}\,+\,\frac{\partial J(0)}{w}\,+\,\ldots\,,\qquad
J(w)J(0)\,\sim\,-\frac{2k}{3 w^2}\,+\ldots\,,
\ee
to immediately deduce that $a=3/(4k)$ and
\be
T_{\rm eff}\,=\, T\,+\frac{3}{4k} J^2\,.
\ee
Inserting the OPE with the effective stress tensor modified by the Sugawara term into the four-point correlator for the R\'enyi entropy, we recover  the short distance expansion \eqref{hol_EE_short} of the holographic EE at large-$c$ in the limit $n\to 1$.

\section{OTO correlator and Lyapunov exponent}
We are now in  position to examine  time dependent questions in this large-$c$ theory using the holographic prescription for higher spin entanglement entropy in the finite temperature CFT undergoing a local quench. The local quench is created by the insertion of a pair of primary operators at the origin at an instant of time. This leads to a time dependent change in the entanglement entropy of a spatial interval when the resulting excitations enter the interval. The holographic dual of this local quench is the backreacted solution for a  black hole (thermal state) with an infalling massive particle, which starts its motion near the boundary and falls towards the black hole horizon. The mass $m$ of the infalling particle  is proportional to the conformal dimension of the operator injecting the quench in the CFT. 

In ordinary gravity the backreacted geometry associated to the infalling particle is obtained by a coordinate transformation and boost on a conical deficit state (see appendix \ref{infallingcoords} for these transformations) \cite{Caputa:2014eta}. This is a so-called shockwave geometry. In higher spin gravity, the notions of metric and spacetime geometry  are gauge dependent. Furthermore, the Wilson line prescription for the entanglement entropy is only sensitive to the location of the endpoints. For this reason, we only need to know the transformation of the endpoints of the Wilson line in a conical deficit background. For a detailed clarification of this point we refer the reader to \cite{David:2017eno}. Here we present the method and the results. 

\subsection{Conical deficit}
In Lorentzian signature, the endpoints of the entangling interval are given by light-cone coordinates $(w_{2}, \bar w_{2})$ and $(w_3, \bar w_3)$  in eq.\eqref{OTOcoords}, while the locations of the operator insertions giving rise to the local quench are at $(w_1,\bar w_1)$ and $(w_4, \bar w_4)$. The operator insertions in imaginary time ($w_1=-i\epsilon$ and $w_4=i\epsilon$) have the effect of spreading out the pulse generated by the quench. The first step is to write down the Wilson line in a conical deficit state:
\be
a\,=\,a_-\,d\xi^-\,,\qquad\qquad \bar a\,=\,\bar a_+\,d\xi^+\,.
\ee
where $a_\mp$ have exactly the same form as $a_w$ and $\bar a_{\bar w}$ respectively, in eqs.\eqref{a} and \eqref{bara}.  For simplicity we take
\be
{\cal G}^+\,=\,{\cal G}^-\,=\,{\cal G}_{\cal O}\,\,,\qquad q\,=\,q_{\cal O}\,,
\ee
as the spin-$\tfrac32$ and $U(1)$ charges of the state.
The key point is that for a conical deficit state we do not impose smoothness conditions on the holonomy. Consequently, the holonomies are nontrivial and encoded in the traces. The energy of the conical deficit with respect to the non-trivial ground state of the CFT is proportional to $\delta_{\cal O}$ and is given by ${\rm Tr}\, a_-^2$\,,
\be
\left(w_{-2}+\frac{3q^2}{4k}\right)\,=\,-\frac{c}{6}\left(1-\frac{\delta_{\cal O}}{c}\right)\,.
\ee
When $\delta_{\cal O}=0$ we recover the nontrivial ground state $|\Phi\rangle$ of the $W_3^{(2)}$ theory with $\Delta_\Phi=-\frac{c}{8}$.
In order to compute the EE using the Wilson line prescription we need the eigenvalues of the connection $a_-$ (and $\bar a_+$).  The eigenvalues $\nu_{1,2,3}$ are the roots of the cubic,
\be
\nu^3\,+\,\nu\left(1-\,\frac{\delta_{\cal O}}{c}\right)\,-\,\left(\pi^2\frac{{\cal G}_{\cal O}^2}{k^2} \,-\,\frac{q_{\cal O}^3}{k^3}\,+\,\frac{q_{\cal O}}{k}\left(\frac{\delta_{\cal O}}{c}\,-\,1\right)\right)\,=\,0\,.
\ee
For the uncharged conical deficit, with ${\cal G}_{\cal O}=q_{\cal O}=0$,
\be
\left(\nu_1,\,\nu_2,\,\nu_3\right)\,=\,\left(-i\sqrt{1-\delta_{\cal O}/c}\quad, \,0\,\,,\,\, i\sqrt{1-\delta_{\cal O}/c}\right)\,.
\ee
Now, we want to consider a situation where the mass and charges of the state are all $\sim O(c)$ in the large-$c$ limit.  We also want these quantities to scale in accordance with dimensional analysis with the infinitesimal width $\epsilon$ of the pulse generating the quench, 
so that the energy and charges of the pulse are fixed in the limit of small width \cite{David:2017eno}.  This means that for a quench of width $\epsilon$, we need to take\footnote{For a holomorphic current $W_s(z)$ with spin-$s$, the charge density in a pulse of width $\epsilon$ behaves as $\langle W_s \rangle \sim q_s/\epsilon^s$ by dimensional analysis, where $q_s$ is a dimensionless charge. The total charge in the pulse is then $\sim q_s/\epsilon^{s-1}$. To keep this fixed in the limit $\epsilon\to 0$, we must scale the charges as $q_s\sim \epsilon^{s-1}$.} 
\be
{\delta}_{\cal O}\,=\, E\,\epsilon\,,\qquad {{\cal G}_{\cal O}}\,=\,\frac{g}{6\pi}\,\epsilon^{1/2}\,,\qquad q_{\cal O}\,=\, \frac{\tilde q}{6}\,.
\ee
For simplicity,  we will set $q_{\cal O}$ to zero below.
Keeping the  leading corrections to the eigenvalues in a small $\epsilon$ expansion we obtain,
\bea
\nu_1\,=\,-i\,+\,\epsilon\,\left(i \frac{E}{2c}\,-\,\frac{g^2}{2c^2}\right)\,,\qquad\nu_2\,=\,\epsilon \,\frac{g^2}{c^2}\,,\qquad \nu_3\,=\,\nu_1^*\,.
\eea
\subsection{Infalling conical defect and EE}
The boost and coordinate transformation \eqref{infallingtransform} which maps the conical deficit configuration into an infalling pulse of width $\epsilon$ in a thermal state with temperature $\beta_\Phi$, acts on the boundary coordinates of the anchor points of the Wilson line as,
\be
e^{i\xi^\pm_{1,2}}\,=\,e^{2\pi i \epsilon/\beta_\Phi}\,\frac{\sinh\tfrac{\pi}{\beta_\Phi}\left(\ell_{1,2}\pm t\mp i\epsilon\right)}{\sinh\tfrac{\pi}{\beta_\Phi}\left(\ell_{1,2}\pm t \pm i\epsilon\right)}\,.
\label{boost}
\ee 
It also has an action on the radial (AdS) coordinate of the endpoints of the Wilson line. 
The thermal state in question has no background one-point functions for ${\cal G }$ and $q$, aside from the charge of the conical deficit itself.
Applying the formula \eqref{WL_holographic_3/2} (replacing $(\Delta w, \Delta\bar w)$ by $\Delta\xi^\pm\,=\,(\xi_2^\pm- \xi_1^\pm)$), the strategy is to first compute  the Wilson line in the conical deficit state and subsequently apply the transformation \eqref{boost}.  Defining the cross-ratio on the thermal cylinder with period $\beta_\Phi$, 
\begin{align}
z = \frac{i \sin \left( \frac{ 2 \pi \epsilon}{\beta_\Phi} \right) \sinh \frac{\pi}{\beta_\Phi}(\ell_2-\ell_1)}{\sinh \frac{\pi}{\beta_\Phi}(\ell_1-t+i\epsilon) \sinh \frac{\pi}{\beta_\Phi}(\ell_2-t-i\epsilon)}\,,
\end{align}
the holographic EE depends on the combinations,
\be
e^{i\Delta\xi^-}\,=\,(1-z)\,, \qquad e^{i\Delta\xi^+}\,=\,(1-\bar z)\,,
\ee
where $\bar z$ is obtained from $z$ by the replacement $t\to -t$.
Then we use the expression for fundamental representation in eq.\eqref{WL_holographic_3/2} to obtain the Wilson line in the adjoint representation, 
\be
W_{\rm Ad}\,=\,W_{\rm fund}\,W_{\overline{\rm fund}}\,,
\ee
keeping in mind the replacements $\Delta w \to \Delta\xi^-$ and $\Delta \bar w\to \Delta\xi^+$. The complete result 
can be rewritten in terms of the cross-ratios  $z$ and $\bar z$ as,
\begin{align} \label{WL_cross-ratio}
{W}_{\rm Ad} = & \frac{16 (z \bar{z})^{-2}}{(\lambda_1 \lambda_2 \lambda_3)^2} \sinh^4 \tfrac{\pi}{\beta_\Phi} (\ell_2-\ell_1) \left[ \sum_{i=1}^3 \lambda_i^{-1} \left( \tfrac{q}{k}+\nu_i \right) \left( \tfrac{q}{k}+\nu_{i+1} \right) \left( \frac{(1-z)^{i \lambda_i}-1}{(1-z)^{\frac{i \lambda_i}{2}-\frac{1}{2}}} \right)^2 \right] \nonumber\\
\times & \left[ \sum_{i=1}^3 \lambda_i^{-1} \left( \tfrac{q}{k}+\nu_i \right) \left( \tfrac{q}{k}+\nu_{i+1} \right) \left( \frac{(1-{\bar z})^{i \lambda_i}-1}{(1 - {\bar z})^{\frac{i \lambda_i}{2}-\frac{1}{2}}} \right)^2 \right],
\end{align}
where $\nu_4\equiv \nu_1$. The holographic EE following the quench
is given by the logarithm of the Wilson line with normalization appropriate for the non-principal embedding,
\be
S_{\rm EE}\,=\,\frac{k}{\sigma_{1/2}}\,\ln W_{\rm Ad}\,,\qquad c=6k\,,\qquad \sigma_{1/2}=2\,.
\ee
Using eq. \eqref{WL_cross-ratio} we see that the EE satisfies the basic requirement that for early and late times ($\ell_{1,2}\gg t$ and $t\gg \ell_{1,2}$) when $|z| \ll 1$, it correctly matches the single interval EE of the $W_3^{(2)}$ theory in the zero charge state \eqref{EE_hol}, up to additive constants independent of the interval length. This matching requires $\beta_\Phi \,=\, \beta/2$ as expected.

At intermediate times when the pulse generated by the quench enters the interval of interest at $t\simeq \ell_1$ and exits at $t\simeq \ell_2$, the entanglement entropy is expected to jump. The jump occurs due to a branch point discontinuity in $S_{\rm EE}$ as a function of $z$ at  $z=1$. When the excitation or pulse is deep inside the interval, say in the limit $\ell_2 \gg t \gg \ell_1$, the cross-ratio completes a clockwise excursion around the branch point:
\be
(1-z)\,\to\,(1-z) e^{-2\pi i}\,.
\ee
The resulting jump in the entanglement entropy can be evaluated as a function of $z$ in the limit $\epsilon \to 0$. We find that for the conical deficit without $U(1)$ charge
\be
\Delta S_{\rm EE}\,=\,\frac{c}{6}\ln\left(1+\frac{\beta_\Phi\, E}{2c}\,{\cal Z}_{\ell_1, \ell_2}^{-1}(t)\right)\,+\,O(\epsilon)\,,
\ee
where
\be
{\cal Z}_{\ell_1, \ell_2}(t)\,=\,\frac{\sinh \frac{\pi}{\beta_\Phi}(\ell_2-\ell_1)}{\sinh \frac{\pi}{\beta_\Phi}(t-\ell_1) \sinh \frac{\pi}{\beta_\Phi}(\ell_2-t)}\,.
\ee
In the limit of large interval length when the excitation is deep inside the interval, the magnitude of the jump in EE becomes
\be
\Delta S_{\rm EE}\big|_{\ell_2\gg t\gg \ell_1}\,=\,\frac{c}{6}\ln\left(1+\frac{\beta_\Phi\, E}{4c}\right)\,.
\ee
As explained in \cite{David:2017eno}, this Wilson line correlator  deep in the quenched  regime can be analytically continued to yield the out of time order (OTO) correlator which exhibits exponential or chaotic growth in the so-called Regge limit. To reach this regime we continue $z$ through the branch cut emanating from $z=1$ to small values of $z$ with ${\rm Im}(z)>0$. This means taking the $t\to\infty$ limit of ${\cal Z}_{\ell_1,\ell_2}(t)$ while staying on the second sheet,
\be
W_{\rm Ad}\big|_{\rm OTO}\,\sim\,\sinh^4 \frac{2\pi}{\beta_\Phi}\left(\ell_2-\ell_1\right) \left(1\,-\,\frac{\beta_\Phi E}{4 c}\,e^{\tfrac{2\pi}{\beta_\Phi}(t-\ell_2)}\right)\,,
\ee
where we have taken $t\gg \ell_2\gg \ell_1$. The exponential growth of this correlator has the associated Lyapunov exponent
\be
\boxed{\lambda_L\,=\,\frac{2\pi}{\beta_\Phi}\,=\,\frac{4\pi}{\beta}}\,.
\ee
This violates the original bound proposed in \cite{mss}. It has been shown in general \cite{Perlmutter:2016pkf} that semiclassical holographic theories with a finite tower of higher spin fields violate the bound of \cite{mss}. In particular, excitations carrying a higher spin $s>2$ have an associated Lyapunov exponent $\lambda_{L}^{(s)} = 2\pi (s-1)/\beta$. Here we see a different mechanism at work, namely the existence of a ground state with negative conformal dimension\footnote{Recently, \cite{deMelloKoch:2019ywq} demonstrated that 
the chaos bound is  violated in the 
fishnet theory which is also known to be nonunitary. 
One of the Lyapunov exponents  evaluated in the rotating BTZ black hole  background also shows 
an apparent violation of the chaos bound \cite{Jahnke:2019gxr, Poojary:2018esz, Stikonas:2018ane}
This  violation has been explained due to the ``effective temperature"
of the left moving modes of the CFT dual to the BTZ black hole.}.

\section{Discussions}
The semiclassical limit of the holographic $W_3^{(2)}$ CFT exhibits various interesting features that are worth exploring further. We have focussed our attention on the vacuum structure of this theory which breaks conformal invariance as a consequence of  nonzero, negative conformal dimension,\footnote{It would be interesting to understand how this picture generalizes to the non-principal embeddings of ${\rm SL}(2)$ in ${\rm SL}(N)$ for $N>3$. Thermodynamics of such non-principal embeddings for the ${\rm SL}(4)$ theory were studied in \cite{Ferlaino:2013vga}. We expect that ground states with negative conformal dimensions will persist, but the precise values of the Lyapunov exponents and their dependence on the embedding would be worthy of further study.} which in turn manifests itself in the violation of the chaos bound of \cite{mss}. 

The large degeneracy of ground states parametrised by expectation values of the $U(1)$ and spin-$\tfrac32$ charges poses nontrivial questions. The zero modes of the spin-$\tfrac32$ currents are like ladder operators and the ground states are therefore not eigenstates of these charges. Instead each semiclassical ground state is like a coherent state for the half-integer spin charge. The situation is thus distinct from theories with $W_N$ symmetry for integer $N$, where it can be shown explicitly that the CFT vacuum block at large-$c$ is computed by a Chern-Simons Wilson line \cite{deBoer:2014sna}.   In the case with half-integer spin fields, the monodromy problem is complicated by (additional) branch cuts due to non-zero condensates for the half-integer charges. It would be interesting to understand the general results for EE that we found from Wilson lines, directly  from large-$c$ CFT arguments in the presence of these condensates.

A tantalizing result that arose from the  holographic study of higher spin black hole phases in \cite{paper1}, was the appearance of a branch of black hole solutions (branch III of \cite{paper1} and in figure \ref{fvt}) which interpolated between the $W_3^{(2)}$ theory in the UV and a holographic $W_3$ CFT in the IR. This branch corresponds to a deformation of the $W_3^{(2)}$ CFT by a chemical potential $\lambda$ for the spin-$\tfrac32$ charges i.e. via introduction of the CFT deformation \eqref{chemical}. Interestingly, this holomorphic (plus antiholomorphic) chemical potential deformation can be studied in conformal perturbation theory from a purely CFT standpoint, following the methods described in \cite{Datta:2014ska}. Up to order $\lambda^2$,
\be
\ln Z\,=\,\ln Z_0 \,-\,\lambda\int d^2 w\, {\cal G}\,\,+\,\frac{\lambda^2}{2}\int d^2 w_1\int d^2 w_2 \frac{3\pi^2\, \langle J\rangle\,(2 g_+\,g_-)}{\beta^2\sinh^2\left(\frac{\pi}{\beta}(w_1-w_2)\right)} \,,
\ee
where we used the ${G^+(w_1) G^-(w_2)}$ OPE  to infer the thermal correlator in the excited state and only the non-vanishing contributions are shown. Importantly, the first non-trivial contributions to the free energy arise from the one-point functions for the spin-$\tfrac32$ current and the $U(1)$ current.
Using the value of the high temperature free energy of the $W_3^{(2)}$ theory discussed in this paper, and the one-point functions for the VEVs of ${\cal G_\pm}$ according to the Casimir conditions with finite-$\lambda$ corrections (appendix \ref{app:lambda}) we obtain
\be
\frac{1}{\tilde L}\ln Z\,=\,\frac{4\pi k}{\beta}\left[1\,+\,\lambda\,\frac{2\sqrt {2\, \beta}}{3^{3/4}\,\sqrt\pi}\,+\,\lambda^2\,\frac{\sqrt 3\,\beta}{\pi}\,\ldots\right]\,,
\ee
which agrees precisely with the partition function corrections found for the branch III black hole solution in \cite{paper1} from Chern-Simons theory. Given that this branch matches onto the $W_3$ theory in the IR, it would be extremely interesting to understand the full expansion in powers of $\lambda$ from CFT and its interpretation from the IR perspective. A large-$c$ CFT understanding of the vacuum block for $W_3^{(2)}$ would help develop chemical potential corrections to EE from the CFT. The lowest order corrections to EE from the Chern-Simons Wilson line prescription have been computed in appendix \ref{app:lambda} and shown to agree with the thermal entropy corrections in the thermodynamic limit.

\acknowledgments
We would like to thank Daniel Gr\"umiller, R. Loganayagam and Matt Headrick for discussions on various aspects of this work. 
 TJH and SPK would like to acknowledge the STFC Consolidated Grant ST/P00055X/1 for support.
SK’s research has been supported by FWO-Vlaanderen (projects G044016N and G006918N) and by Vrije Universiteit Brussel through the Strategic Research Program “High-Energy Physics”.

\appendix
\section{Commutation relations for $W_3^{(2)}$}
The modes of the $W_3^{(2)}$ currents satisfy the following commutation relations which can be deduced easily from the OPEs:
\bea
&&\left[ {L}_n, {L}_m \right] \, = \,(n-m)\, L_{n+m}\,+\,\tfrac{c}{12}\,{n(n^2-1)}\delta_{n+m,0}\,, \qquad \left[ {L}_n, {J}_m \right]\,=\,-m\, J_{n+m} \,,\nonumber\\\nonumber\\\nonumber
&&\left[ {L}_n, {G}_m^\pm \right]\,=\,\left(\tfrac{n}{2}-m\right) G_{n+m}^\pm \,,\qquad
\left[ J_n,G_m^{\pm} \right] \, =\, \pm G_{n+m}^{\pm}\,,\qquad
\left[ J_n,J_m \right] \, =\, - \,\tfrac{2k-3}{3}\,n\,\delta_{n+m,0}\,,\nonumber\\
\label{modexp}\\\nonumber
&&\left[ G_n^+, G_m^- \right] \, =\, \tfrac{1}{2}(k-1)(2k-3) \left( n^2 -\tfrac{1}{4} \right) \delta_{n+m,0} + (k-3)\hat{L}_{n+m} -\tfrac{3}{2}(k-1)(n-m) J_{n+m}\, \nonumber
\\\nonumber\\\nonumber
&&\qquad\qquad +\,3\sum_\ell :J_{n+m-\ell} J_\ell :\,.
\eea

\section{Generators in non-principal embedding}
For the non-principal embedding of ${\mathfrak {sl}}(2, {\mathbb R})$ in ${\mathfrak{sl}}(3,{\mathbb R})$, we define the generators
\begin{align}
\hat{L}_0 & = \frac{1}{2} L_0, \qquad \hat{L}_\pm = \pm \frac{1}{4} W_{\pm 2}, \qquad J_0 = \frac{1}{2}W_0,\label{generators}\\
G^\pm_{1/2} & = \frac{1}{\sqrt{8}}\left( W_1 \mp \ell_1 \right), \qquad G^\pm_{-1/2} = \frac{1}{\sqrt{8}} \left(L_{-1} \pm W_{-1} \right)\,,\nonumber
\end{align}
where the ${\mathfrak{sl}}(3, {\mathbb R})$ generators satisfy,
\bea
&&[L_i, L_j]\,=\,(i-j)\,L_{i+j}\,,\qquad [L_i, W_m]\,=\,(2i-m)\,W_{i+m}\,,\\\nonumber
\\\nonumber
&&[W_m, W_n]\,=\,-\frac{1}{3}(m-n)(2m^2 +2n^2-mn-8)\,L_{m+n}\,,
\eea
with $i,j=-1,0,1$, and $m,n\,=\,-2,1,0,1,2$.
\section{Infalling conical deficit}
\label{infallingcoords}
We summarize the map which takes a conical deficit state in AdS$_3$ to an infalling particle in a black hole background. The transformation consists of a boost with parameter $\tilde \epsilon$. The locations of the endpoints of the Wilson line transform according to:
\bea
&&e^{\rho_{P,Q}}\,=\label{infallingtransform}\\\nonumber
&&\frac{\Lambda R\beta}{2\pi}\sqrt{\sinh^2\left(\frac{2\pi x_{P,Q}}{\beta_\Phi}\right)\,+\,\left(\frac{\beta_\Phi}{2\pi\tilde\epsilon}\cosh\left(\frac{2\pi t}{\beta_\Phi}\right)\,-\,\sqrt{\left(\tfrac{\beta_\Phi}{2\pi\tilde\epsilon}\right)^2-1}\cosh\left(\frac{2\pi x_{P,Q}}{\beta_\Phi}\right)\right)^2}
\\\nonumber\\\nonumber
&&\tan(\tau_{P,Q})\,=\,\frac{2\pi\tilde\epsilon}{\beta_\Phi}\frac{\sinh\left(\frac{2\pi t}{\beta_\Phi}\right)}{\cosh\left(\frac{2\pi x_{P,Q}}{\beta_\Phi}\right)\,-\,\sqrt{1-\left(\tfrac{2\pi\tilde\epsilon}{\beta_\Phi}\right)^2}\cosh\left(\frac{2\pi t}{\beta_\Phi}\right)}
\\\nonumber\\\nonumber
&&\tan(\phi_{P,Q})\,=\,\frac{2\pi\tilde\epsilon}{\beta_\Phi}\frac{\sinh\left(\frac{2\pi x_{P,Q}}{\beta_\Phi}\right)}{\cosh\left(\frac{2\pi t}{\beta_\Phi}\right)\,-\,\sqrt{1-\left(\tfrac{2\pi\tilde\epsilon}{\beta_\Phi}\right)^2}\cosh\left(\frac{2\pi x_{P,Q}}{\beta_\Phi}\right)}
\eea
Here $\Lambda$ is a UV cutoff representing the AdS boundary. The spatial endpoints of the entangling interval are at $x_P=\ell_1$ and $x_Q=\ell_2$. The boost parameter $\tilde\epsilon$ is related to the width of the pulse produced by the local quench:
\be
\frac{2\pi\tilde\epsilon}{\beta_\Phi}\,=\,\sin\frac{2\pi\epsilon}{\beta_\Phi}\,.
\ee
 
\section{Chemical potential corrections to entanglement entropy}
\label{app:lambda}
Consider the C-S connection representing a  ${ W}^{(2)}_3$ black hole where the chemical potential, $\lambda$, is nonzero and both spin-$3/2$ and $U(1)$ currents are present switched on, 
\begin{align}
a & = \left(  \frac{1}{4} W_2 +\frac{3q}{4k}  W_0 + \frac{w_{-2}}{4k} W_{-2} + \frac{\pi{\cal G}}{2k} L_{-1} \right) dx^+ + \lambda \left( L_1 +\frac{3q}{2k}L_{-1}-\frac{\pi {\cal G}}{4k} W_{-2}\right) dx^-\\
\bar{a} & = \lambda \left( L_{-1}+\frac{3q}{2k}\ell_1 -\frac{\pi{\cal G}}{4k} W_2 \right) dx^+ - \left( -\frac{1}{4} W_{-2}-\frac{3q}{4k}W_0-\frac{w_{-2}}{4k}W_2 -\frac{\pi{\cal G}}{2k}\ell_1 \right)dx^-.\nonumber
\end{align}
The requirement of flatness implies
\begin{align}
w_{-2} = \frac{9}{4k} q^2,
\end{align}
and the holonomy conditions for a smooth black hole solutions are,
\begin{align}
& \frac{4\pi^2k}{3\beta^2} = \frac{q^2}{k}-\lambda \pi {\cal G} -2\lambda^2 q\\
& \pi^2 {\cal G}^2 + 4\frac{q^3}{k}+6\pi \lambda {\cal G} q +24 \lambda^2 q^2 - 4\pi k \lambda^3 {\cal G} =0. \nonumber
\end{align}
There are four sets of roots for the charges $q$ and ${\cal G}$, but only two of these are real. They correspond to branches III and IV identified in \cite{paper1}. For branch III which was identified as the thermodynamically stable one, $q$ and ${\cal G}$ can be expanded as a power series in $\lambda$:
\begin{align} \label{holonomy_soln_beta}
q & = -\frac{2 \pi  k}{\sqrt{3} \beta } +\frac{\sqrt{2 \pi } k \lambda }{\sqrt[4]{3} \sqrt{\beta }}-\frac{1}{2} 3 k \lambda ^2 + \dots\\
{\cal G} & = -\frac{4 \sqrt{2 \pi } k}{3^{3/4} \beta ^{3/2}}+\frac{4 \sqrt{3} k \lambda }{\beta }-\frac{3\ 3^{3/4} k \lambda ^2}{\sqrt{2 \pi } \sqrt{\beta }} + \dots .\nonumber
\end{align}
We have mostly limited our discussion to the $\lambda\to 0$ limit in this paper.

\subsection{Thermodynamics}
Consider the thermodynamic action,
\begin{align}
I_{\text{th}} = I_{\text{on-shell}} -\pi {\cal G} \lambda \beta - \frac{4\pi^2 k}{\beta}
\end{align}
where,
\begin{align}
I_{\text{on-shell}} = - \beta \left( w_{-2} + \frac{3}{4k}q^2 +6 \lambda^2 q \right)
\end{align}
The partition function is,
\begin{align}
\frac{1}{\tilde L}\ln Z = - I_{\text{th}} = \beta \left( w_{-2} + \frac{3}{4k}q^2 +6 \lambda^2 q + \pi {\cal G} \lambda+  \frac{4\pi^2 k}{\beta^2} \right)\,,
\end{align} 
where $\tilde L$ is the spatial extent of the CFT.
The thermal entropy is obtained by taking derivatives with respect to the temperature $T=\beta^{-1}$,
\begin{align}
S = \partial_T \left( \frac{T}{L} \ln Z \right) = \frac{6q}{k} \partial_T q + 6\lambda^2 q +\pi \lambda \partial_T {\cal G} + \frac{8\pi k}{\beta}
\end{align}
which gives the following expression upon using using the one-point functions,
\begin{align}
S & = \frac{L}{2\pi} \beta \left( 4 \left( w_{-2}+\frac{3}{4k}q^2 \right) -6 \pi \lambda {\cal G} \right)\\
& =  \frac{4c  \pi L}{3 \beta} \left( 1- \lambda \frac{3^{1/4} \sqrt{\beta}}{\sqrt{2\pi}} + \lambda^2 \frac{\sqrt{3} \beta}{2\pi} + \dots \right)\,,\nonumber
\end{align}
\subsection{Holographic entanglement entropy with $\lambda\neq 0$}
In the paper we have confined ourselves mostly to the situation with vanishing chemical potential. Here we quote some results for the single interval entanglement entropy computed using the holomorphic prescription \cite{deBoer:2013vca, paper1}, as a power series in the spin -$3/2$ chemical potential $\lambda$,
\bea\label{EE_lambda}
&&\exp\left( \frac{6}{c}{S_{\rm EE}}\right)  = \frac{\beta ^4}{9 \pi ^4} \sinh ^4\left(\tfrac{\pi  L }{ \beta }\right) \left(\cosh \left(\tfrac{2 \pi  L }{\beta }\right)+2\right)^2\,+\,\\\nonumber\\
&&+ \,\lambda\frac{2 \sqrt{2} \beta ^{7/2} }{3\ 3^{3/4} \pi ^{9/2}}  \sinh ^5\left(\tfrac{\pi  L }{ \beta }\right) \left(\cosh \left(\tfrac{2\pi  L }{\beta }\right)+2\right) \left(\beta  \sinh \left(\tfrac{\pi  L }{ \beta }\right)-4 \pi  L  \cosh \left(\tfrac{\pi  L }{ \beta }\right)\right) \nonumber\\\nonumber\\
&& +\frac{\beta ^3 \lambda ^2 }{12 \sqrt{3} \pi ^5} \sinh ^4\left(\tfrac{\pi  L }{ \beta }\right) \left[-4 \left(\beta ^2-20 \pi ^2 L^2\right) \cosh \left(\tfrac{2\pi  L }{\beta }\right)+3 \beta ^2 +32 \pi ^2 L ^2 \right. \nonumber\\
&& \left. + \left(\beta ^2+8 \pi ^2 (2L)^2\right) \cosh \left(\tfrac{4 \pi  L }{\beta }\right) +16 \pi  \beta  L  \sinh \left(\tfrac{2\pi  L }{\beta }\right)-8 \pi  \beta  L  \sinh \left(\tfrac{4 \pi  L }{\beta }\right)\right]\nonumber\,.
\eea
In the limit of large interval length $L /\beta\gg 1$, the expansion in $\lambda$ agrees with the thermal entropy above,
\begin{align} \label{EE_large_int}
 S_{\rm EE}\,\big|_{L/\beta\gg 1}\,=\, \frac{4c \pi L}{3 \beta} \left( 1- \lambda \frac{3^{1/4} \sqrt{\beta}}{\sqrt{2\pi}} + \lambda^2 \frac{\sqrt{3} \beta}{2\pi} + \dots \right).
\end{align}

\subsection{$\lambda$ corrections to ground state expectation from CFT}
In this section we evaluate the $\lambda$ corrections to the ground state expectation value of the stress tensor. We evaluate the following,
\begin{align} \label{Tgslambda}
\frac{\langle \Phi(\infty) T(x) \exp\left( \lambda \int d^2y \; \psi^+(y) \right) \Phi(-\infty) \rangle}{\langle \Phi(\infty) \exp\left( \lambda \int d^2y\; \psi^+(y) \right) \Phi(-\infty) \rangle}= \langle T(x) \rangle_{\rm \Phi} + \lambda \left( \frac{ N^{(1)} - D^{(1)} \langle T(x) \rangle_{\rm\Phi} }{\langle \Phi(\infty)  \Phi(-\infty) \rangle}\right)\\
+\frac{\lambda^2}{2} \left( \frac{N^{(2)} - D^{(2)}\langle T(x) \rangle_{\rm \Phi}+\frac{2\left(D^{(1)}\right)^2\langle T(x) \rangle_{\rm \Phi}}{\langle \Phi(\infty)  \Phi(-\infty) \rangle} - \frac{2N^{(1)} D^{(1)}}{\langle \Phi(\infty)  \Phi(-\infty) \rangle}}{\langle \Phi(\infty)  \Phi(-\infty) \rangle } \right) + \dots\nonumber
\end{align}
where,
\begin{align}
N^{(1)} & = \int d^2y \; \langle \Phi^\dagger(\infty) T(x) \psi^+(y) \Phi(-\infty)\rangle\\
N^{(2)} & = \int d^2y_1 \int d^2y_2 \; \langle \Phi^\dagger(\infty) T(x) \psi^+(y_1) \psi^+(y_2) \Phi(-\infty)\rangle \nonumber\\
D^{(1)} & = \int d^2y \; \langle \Phi^\dagger(\infty) \psi^+(y) \Phi(-\infty)\rangle \nonumber\\
D^{(2)} & = \int d^2y_1 \int d^2y_2 \; \langle \Phi^\dagger(\infty) \psi^+(y_1) \psi^+(y_2) \Phi(-\infty).\rangle \nonumber
\end{align}
All the integrals over $y$ run from $-\infty \; \text{to} \; \infty$, and
\begin{align}
\psi^+(y) = g_+ G^+(y)+g_- G^-(y).
\end{align}
Using,
\begin{align}
D^{(1)} & = {\cal G} \int d^2 y\\
N^{(1)} & = {\cal G} \left[ \left( \frac{4\pi^2}{\beta^2} \Delta_\Phi -\frac{c\pi^2}{6\beta^2} \right) \int d^2y + \int d^2 y \frac{3\pi^2}{2\beta^2} \frac{1}{\sinh^2\left(\frac{\pi y_1}{\beta}\right)} \right]\nonumber.
\end{align}
Substituting the above equations in the ${O}(\lambda)$ term of equation \eqref{Tgslambda}, the first two terms of $N^{(1)}$ cancel with $D^{(1)} \langle T \rangle_{\rm \Phi}$ by using equation the expectation value of the stress tensor in the excited thermal state at $\lambda=0$ \eqref{stress1pt} . Therefore we get,
\begin{align}
\langle T(x) \rangle_{\rm gs} = - \frac{4 k \pi^2}{\beta^2} + \lambda \frac{4 \sqrt{2}\, 3^{1/4}k\pi^{3/2}}{\beta^{3/2}} + \dots
\end{align}
which agrees with the holographic answer to this order.

%\appendix

%\bibliographystyle{JHEP}
%\bibliography{w23}

\end{document}